%% file: ms.tex
\documentclass[12pt,preprint]{aastex}

\usepackage{natbib}
\slugcomment{version 01/17/2002}
\shorttitle{D, N, O towards HZ 43A}
\shortauthors{Kruk et al.}

\newcommand{\LA}{Lyman-$\alpha$}
\newcommand{\LB}{Lyman-$\beta$}
\newcommand{\LG}{Lyman-$\gamma$}
\newcommand{\HZ}{HZ\,43A}
\newcommand{\fuse}{{\it FUSE}}

\newcommand{\euve}{{\it EUVE\/}}
\newcommand{\orfeus}{{\it ORFEUS\/}}
\newcommand{\hst}{{\it HST\/}}

\newcommand{\ghrs}{GHRS}
\newcommand{\Hmol}{\mbox{H$_{2}$}}

\newcommand{\chisq}{\ensuremath{\chi^2}}
\newcommand{\err}[2]{\ensuremath{^{+ #1}_{- #2}}}
\newcommand{\kms}{km s\ensuremath{^{-1}}}
\newcommand{\cmsq}{cm\ensuremath{^{-2}}}

\newcommand{\Nh}{\ensuremath{\:{\rm N_{H\:I}}}}
\newcommand{\Hone}{H\,{\sc i}}
\newcommand{\Htwo}{H\,{\sc ii}}
\newcommand{\Done}{D\,{\sc i}}
\newcommand{\Arone}{Ar\,{\sc i}}

\newcommand{\Ctwo}{C\,{\sc ii}}
\newcommand{\Cthree}{C\,{\sc iii}}
\newcommand{\Heone}{He\,{\sc i}}
\newcommand{\Hetwo}{He\,{\sc ii}}
\newcommand{\Hethree}{He\,{\sc iii}}
\newcommand{\Oone}{O\,{\sc i}}

\newcommand{\Osix}{O\,{\sc vi}}
\newcommand{\None}{N\,{\sc i}}
\newcommand{\Ntwo}{N\,{\sc ii}}
\newcommand{\Nthree}{N\,{\sc iii}}
\newcommand{\Sitwo}{Si\,{\sc ii}}
\newcommand{\Mgtwo}{Mg\,{\sc ii}}
\newcommand{\Fetwo}{Fe\,{\sc ii}}
\newcommand{\lognh}{\ensuremath{\log\:N_{H\,I}}}
\newcommand{\lognd}{\ensuremath{\log\:N_{D\,I}}}
\newcommand{\lognn}{\ensuremath{\log\:N_{N\,I}}}
\newcommand{\lognntwo}{\ensuremath{\log\:N_{N\,II}}}

\newcommand{\logno}{\ensuremath{\log\:N_{O\,I}}}
\newcommand{\vphot}{\ensuremath{v_{ph}}}
\newcommand{\vnt}{\ensuremath{b_{nt}}}
\newcommand{\onesig}{\ensuremath{1\:\sigma}}
\newcommand{\twosig}{\ensuremath{2\:\sigma}}
\newcommand{\threesig}{\ensuremath{3\:\sigma}}

\newcommand{\Teff}{$\:{\rm T_{eff}}$}
\newcommand{\logg}{log\,g}

\begin{document}

\title{Abundances of Deuterium, Nitrogen, and Oxygen toward HZ\,43A: Results 
from the FUSE Mission\altaffilmark{1}}
\altaffiltext{1}{Based on observations made 
with the NASA-CNES-CSA Far Ultraviolet Spectroscopic Explorer.
\fuse\ is operated for NASA by the Johns Hopkins University under NASA
contract NAS5-32985.}

\author{J. W. Kruk\altaffilmark{2}, J. C. Howk, M. Andre, H. W. Moos, 
W. R. Oegerle\altaffilmark{3}, C. Oliveira, K.R. Sembach, 
P. Chayer\altaffilmark{4}}
\affil{Department of Physics and Astronomy, Johns Hopkins University,
Baltimore, MD  21218 USA}

\author{J. L. Linsky, B. E. Wood}
\affil{JILA, University of Colorado and NIST, Boulder, CO, 80309-0440 USA}

\author{R. Ferlet, G. H\'{e}brard, M. Lemoine, A. Vidal-Madjar}
\affil{Institut d'Astrophysique de Paris, 98$^bis$ Boulevard Arago, F-75014,
Paris, France}

\and
\author{G. Sonneborn}
\affil{Laboratory for Astronomy and Solar Physics, NASA/GSFC, Code 681,
Greenbelt, MD 20771  USA}

\altaffiltext{2}{kruk@pha.jhu.edu}
\altaffiltext{3}{Present address:  Laboratory for Astronomy and Solar Physics, 
NASA/GSFC, Code 681, Greenbelt, MD 20771  USA}
\altaffiltext{4}{Department of Physics and Astronomy, University of Victoria,
P.O. Box 3055, Victoria, BC V8W 3P6, Canada}

\begin{abstract}
We present an analysis of interstellar absorption along the line of sight
to the nearby white dwarf star \HZ.  
The distance to this star is 68$\pm$13 pc, and the line of sight extends
toward the north Galactic pole. 
Column densities of \Oone,
\None, and \Ntwo\ were derived from spectra obtained by the Far Ultraviolet
Spectroscopic Explorer (\fuse), the column density of \Done\ was derived
from a combination of our \fuse\ spectra and an archival \hst\ GHRS spectrum,
and the column density of \Hone\ was derived from a combination of the
GHRS spectrum and values derived from \euve\ data obtained from the literature.
We find the following 
abundance ratios (with \twosig\ uncertainties):  
\Done/\Hone\ = $(1.66 \pm 0.28)\times 10^{-5}$, 
\Oone/\Hone\ = $(3.63 \pm 0.84)\times 10^{-4}$, and
\None/\Hone\ = $(3.80 \pm 0.74)\times 10^{-5}$.
The \Ntwo\ column density was slightly greater than that of \None,
indicating that ionization corrections are important when deriving
nitrogen abundances.
Other interstellar species detected along the line of sight 
were \Ctwo, \Cthree, \Osix,
\Sitwo, \Arone, \Mgtwo, and \Fetwo; an upper limit was determined for
\Nthree.  No elements other than \Hone\ were detected in the stellar 
photosphere.
\end{abstract}

\keywords{ISM: abundances---ISM: clouds---stars: individual (\HZ)---stars:
white dwarfs}

\section{Introduction}

Deuterium is one of the primary products of big bang nucleosynthesis
(BBN), and its primordial abundance provides a sensitive measure of
the baryon density of the universe \citep{Schramm:1998, Burles:2000}.
However, deuterium is consumed in stars far more quickly than it is
produced by the P-P cycle or by any other known stellar nucleosynthetic
process \citep{Epstein:1976}; thus, measurements of the deuterium
abundance at the present epoch can only provide lower limits to the
primordial abundance.  Measurements obtained at a variety of redshifts
will sample the chemical evolution of the universe, and may permit
extrapolation to a value for the primordial abundance of deuterium.
Measurements of the abundance of deuterium and various products of
stellar nucleosynthesis, such as oxygen, within our Galaxy will
improve our understanding of chemical evolution as well as provide a
lower limit to the primordial deuterium abundance.  Recent reviews of
deuterium abundance measurements can be found in
\citet{Lemoine:1999}, \citet{Linsky:1998}, and \citet{Moos:2001}.

\citet{Linsky:1998} reviews measurements of the deuterium
abundance for material in the local interstellar cloud (LIC), which
are consistent with a common value for D/H of $(1.5\,\pm\,0.1)
\times\,10^{-5}$.  The LIC and other nearby clouds are thought to be 
embedded in a large bubble of hot, low density gas
($T\approx\,10^{6}K,\ n\approx\,0.005\,cm^{-3}$) that was probably
produced by supernovae and stellar winds arising in the
Scorpius-Centaurus OB association
\citep{Cox:1987, Frisch:1995}.  If this picture is correct, then the
local interstellar medium (LISM) may contain an inhomogeneous mixture
of clumps of older material swept up by the expanding bubble along
with the more-recently processed material in the bubble itself.  Thus,
even in this local environment one may find regions of gas with
somewhat different evolutionary histories and different compositions.
The same processes have been at work throughout the history of the
Galaxy; hence measurements of deuterium and other abundances along
numerous lines of sight will help determine not only the chemical
evolution of the Galaxy, but also the relative timescales for chemical
evolution and the mixing of material.

In this paper we present an analysis of the line of sight to the nearby
white dwarf \HZ.  Companion papers will present similar analyses of the
lines of sight to G\,191-B2B \citep{Lemoine:2001}, WD\,0621-376
\citep{Lehner:2001}, WD\,1634-573 \citep{Wood:2001}, WD\,2211-495
\citep{Hebrard:2001}, BD\,$+28^{\degr}~4211$ \citep{Sonneborn:2001},
and Feige\,110 \citep{Friedman:2001}. An overview will be provided by
\citet{Moos:2001}.

High-resolution spectra of the DA white dwarf \HZ, covering the far
ultraviolet (FUV) wavelength range 905--1187\AA\ were obtained with
the {\it Far Ultraviolet Spectroscopic Explorer} (\fuse) for the
purpose of studying the deuterium abundance of the local interstellar
medium (ISM).  This line of sight is promising for a study of D/H for
several reasons: the star itself is moderately bright, exhibits a
nearly featureless continuum, and has been accurately modelled; the
absorbing interstellar gas appears to have only a single velocity
component and the column density is very low, so many absorption lines
that are ordinarily on the flat part of the curve of growth are either
unsaturated or exhibit only mild saturation effects.  In addition, the
\Hone\ column density along the line of sight can be determined by two
independent methods: from the shape of the Lyman continuum observed by
the {\it Extreme Ultraviolet Explorer} (\euve), and from the \LA\
profile observed with the Goddard High Resolution Spectrograph (GHRS)
onboard the Hubble Space Telescope (\hst).  Systematic uncertainties
in determination of \Hone\ column densities are often the limiting
factor in measurements of the D/H ratio, so the availability of
multiple independent measurements of N(H) is quite valuable.

In Section 2 we discuss the line of sight to \HZ\ and the properties
of the star; in Section 3 we describe the observations and data
reduction procedures; in Section 4 we describe our analysis
procedures, our methods for minimizing systematic errors, and our
measured column densities; and in Section 5 we summarize our results.

\section{Line of Sight and Stellar Properties}
\label{sec_los}
The hot DA white dwarf HZ\,43A (WD\,1314+293) is a member of a
non-interacting binary system with a dMe star.  The cool secondary
star has no significant UV flux.  The coordinates of the star and some
other basic characteristics are given in Table \ref{tab-tgtinfo}.  The
Galactic coordinates, $l = 54.10\degr, b = +84.16\degr$, are close to
the north Galactic pole and the line of sight exits the LIC after
traversing a very short distance: $\sim$0.1pc \citep{Redfield:2000}.
The mean \Hone\ density in the LIC is 0.1\,cm$^{-3}$
\citep{Linsky:2000}, so the corresponding column density is only
$\approx\, 3.1\times10^{16}\,{\rm cm}^{-2}$.  The velocity of the LIC
projected onto the \HZ\ line of sight is -8.9\,\kms\ (all velocities
given in this paper are heliocentric).  The line of sight may
intersect the ``G'' cloud that lies between us and the Galactic
center, and the ``NGP'' cloud that lies between us and the North
Galactic Pole (see \citealt{Sfeir:1999} and the review by
\citealt{Frisch:1995} for a discussion
of the geometry of the nearby clouds lying within the local bubble).
The boundaries of the clouds are not known in detail, but the identity
of each absorber can be determined by its relative velocity.  In particular,
the velocity of the G cloud projected onto the \HZ\ line of sight is
-12.1\,\kms.  The distance to \HZ\ has been determined by several
methods to be approximately 65pc.  \citet{Vennes:1997} obtained 67pc
from a comparison of its apparent and absolute V magnitude (derived
from a model atmosphere), while \citet{Dupuis:1998} obtained a
somewhat smaller value, 56-61pc using a similar method.  Ground-based
parallax measurements have given 48-91pc \citep{Margon:1976}, 53-83pc
\citep{Dahn:1982}, and 55-81pc \citep{vanAltena:1995}.  The only 
discrepant measurement, 25 - 44pc, is from the Hipparcos catalog
\citep{Clark:1999}; the origin of this discrepancy is not understood.
The apparent line-of-sight velocity of the star (including gravitational
redshift) was measured by \citep{Reid:1996} to be +20.6\,\kms.

The atmospheric parameters of \HZ\ are difficult to measure at visible
wavelengths, because of the presence of the bright M star companion only
$\sim 3$\arcsec\ away.  \citet[N93 hereafter]{Napiwotzki:1993} and 
\citet[FKB hereafter]{Finley:1997} obtained optical spectra of
both components and corrected the observed flux of the white dwarf by
subtracting the contribution from the M star.  N93 obtained
$T_{\rm{eff}} = 49,000$ K and $\log g = 7.7$ using NLTE models, while
FKB determined $T_{\rm{eff}} = 50,800$ K and $\log g = 7.99$ using LTE
models.  According to \citet{Napiwotzki:1999}, the
difference between both studies may illustrate the correction in
$T_{\rm{eff}}$ and $\log g$ that one must apply to transform the
atmospheric parameters determined by LTE models to NLTE models.
However, if we apply the \citet{Napiwotzki:1999} correction
($\Delta\log g = -0.13$) to FKB's value (7.99), we obtain a larger
gravity than N93 (7.7).  

Other spectral bands have been used to evaluate the effective
temperature and gravity of \HZ.  For example, \citet{Holberg:1986}
fitted the {\it IUE} \LA\ line profile, obtaining a larger temperature
and gravity than the optical observations performed by N93 and FKB.
Both \citet{Vennes:1992a} and FKB indicated that the visual magnitude
estimated by \citet{Holberg:1986} was too faint.  The visual magnitude
is quite difficult to determine from the ground because of
contamination by the companion star.  If \citet{Holberg:1986} had used
the magnitude determined from \hst\ data by \citet{Bohlin:1995}, the
effective temperature would have been $T_{\rm{eff}} = 50,850$ K, very
close to that determined later by FKB.  The complete Lyman series of
\HZ\ was analyzed by \citet{Dupuis:1998} using {\it ORFEUS} and by
\citet{Kruk:1997} with the {\it Astro-1 Hopkins Ultraviolet
Telescope}, while the EUV wavelength range was studied by
\citet{Dupuis:1995} and \citet{Barstow:1995} using \euve.  If we take
the average of the effective temperature and gravity for all
measurements made from EUV, FUV, and optical observations, we obtain
$T_{\rm{eff}} = 50,300$ K and $\log g = 7.88$, with standard
deviations of 700~K and 0.15~dex (see, e.g., Table 1 of
\citealt{Dupuis:1998}).

For the purpose of this study,
we adopted the following atmospheric parameters:  $T_{\rm{eff}} = 50,900$ K,  
and $\log g = 8.0$.  These are the pre-publication values of FKB
used by \citet{Kruk:1999} 
in the final {\it Astro-2} calibration of the {\it Hopkins
Ultraviolet Telescope}, and used presently in the \fuse\ calibration
(the present \fuse\ flux calibration is defined by \HZ, G\,191$-$B2B, and
GD\,246).

\HZ\ is unusual in that its photosphere appears to
consist entirely of hydrogen, though stars
with similar atmospheric parameters show traces of heavy elements.
\citet{Barstow:1995} set an upper limit
on the helium abundance of $10^{-7}$, based on the \euve\ spectrum of
this star, and \citet{Dupuis:1998} used an \orfeus\ spectrum to
set abundance limits of $10^{-7}$ for S,
$10^{-8}$ for C and N, and $10^{-8.5}$ for Si, P, and Cl.
\citet{Vennes:1991}, N93, and \citet{Dupuis:1995}
came to the same conclusion, though with somewhat higher upper limits.
Therefore, photospheric models for this star employing a pure hydrogen
composition should give accurate results.

\section{Observations}
\subsection{\fuse\ Observations}

\fuse\ comprises four independent co-aligned telescopes and spectrographs.
Two of the channels contain optics coated with aluminum and overcoated
with LiF, and the other two channels contain optics coated with SiC.
Together the four channels span the wavelength range 905-1187\AA.
Most wavelengths in this band are sampled by at least 2 channels, and
roughly one-third of the bandpass is sampled by 3 or 4 channels
(1000\AA\  $<~\lambda~<~$1080\AA).
Comparison of features in the spectra obtained in different channels
can thus be used to test for the effects of fixed-pattern noise, backgrounds, or
other instrumental characteristics that vary from one channel to another.
Each spectrograph has 3 entrance apertures:  LWRS (30\arcsec $\times$ 
30\arcsec), MDRS (4\arcsec $\times$ 20\arcsec), and HIRS (1.25\arcsec $\times$
20\arcsec).  Each aperture illuminates different regions of the detector
and produces spectra with slightly different line-spread functions (LSFs).
Thus, if a source is observed through multiple apertures one can perform
additional tests for the presence of fixed-pattern noise introduced
by the detector,
and one can determine if contamination by geocoronal emission is significant.
Further information on the \fuse\ satellite and instrument on-orbit
characteristics 
are provided by \citet{Moos:2000} and \citet{Sahnow:2000}.

Three \fuse\ observations of \HZ\ have been obtained, two in the LWRS
aperture and one in the MDRS aperture.  Basic information for the
datasets is given in Table \ref{tab-fuseobs}.  The data were obtained
in ``histogram'' mode, in which a two-dimensional spectral image is
accumulated and downlinked.  The LWRS observations will be described
here for the sake of completeness: the signal-to-noise of the LWRS
observations was too low in comparison with the MDRS data for them to
significantly constrain the results.  They were useful primarily for
tests of fixed-pattern noise effects.  Exposure durations for the LWRS
observations ranged from just under 500\,s to just over 800\,s.  The
exposures in the MDRS observation were all approximately 485\,s, apart
from a few exposures that were shortened to roughly 200\,s by delayed
target acquisitions.  The MDRS observation was split into 4 intervals
of approximately equal exposure time, each with a different X-position
of the Focal Plane Assemblies (FPAs).  The shift in FPA position
causes a corresponding shift in the position of the spectrum on the
detector along the dispersion direction.  The 4 positions were at X =
-57\,$\mu$m, 0\,$\mu$m, +33\,$\mu$m, and +125\,$\mu$m with respect to
the nominal position, corresponding to spectral shifts of roughly -12,
0, +5, and +19 pixels.  These shifts span a range several times larger
than the width of typical interstellar absorption lines, and several
times larger than the width of the high-frequency components of the
detector fixed-pattern noise.  This so-called ``focal-plane split''
(``FP-split'') procedure results in a considerable improvement in the
signal-to-noise ratio achieved, even if the the spectra from the
different FPA positions are simply shifted and added with no attempt
to explicitly solve for the detector fixed-pattern noise.

The data were processed using the most recent version of the \fuse\
calibration pipeline, CALFUSE version 1.8.7.  Because there is no
information available for the arrival time of individual photons in
histogram mode, the pipeline applies exposure-averaged corrections for
the Doppler shift induced by the spacecraft orbital motion and for the
rotation of the gratings on orbital timescales.  The primary mirrors
also rotate slightly on orbital timescales, which can lead to
significant changes in the position of the image of the source at the
spectrograph entrance aperture.  As a result, the effective spectral
resolution can vary from one exposure to another and the zero-point of
the wavelength scale varies from one exposure to another.

The rotations of the primary mirrors during an exposure can cause the
stellar image to fall outside of the corresponding spectrograph aperture, 
resulting in loss of signal in that channel for all or part of an exposure.
The Fine Error Sensor presently used to provide pointing information to
the satellite attitude control system is part of the optical path in
the LiF1 channel; hence the target is always maintained at a fixed 
position in the LiF1 spectrograph aperture.  The alignment of the other
channels is usually maintained so that the rotations of the primary
mirrors occurring on orbital timescales do not shift the source image at
each focal plane out of the corresponding 
LWRS aperture.  For the M1010501 observation, no flux was lost in any of
exposures; for P1042301 no flux was lost except that the SiC1
channel fluxes were low by 7\%, 10\%, and 19\% in exposures 13, 16,
and 20, respectively.  The SiC1 spectra obtained in these exposures
were scaled accordingly before being combined with the other exposures.
For MDRS observations, peakups are performed twice in each orbit in
order to try to correct for the mirror rotations by moving the spectrograph
apertures (FPA)s in the dispersion direction
to maximize the signal prior to the start of an exposure.  
For the P1042302 MDRS observation, this procedure
was fairly successful, providing effective exposure times of 82\% for LiF2,
62\% for SiC1, and 76\% for SiC2.  The necessity of moving the FPAs
to compensate for rotations of the primary mirrors means that there
is actually a distribution of FPA positions clustered about the mean offsets
that define the FP-split sequence described above.

The spectra obtained in different exposures generally need to be co-aligned
(shifted to a common wavelength zero-point) prior to combining.
This is caused in part by the lack of corrections for primary mirror
rotations in the present pipeline, and in part by shortcomings in
the present corrections for grating rotations.  For \HZ\ there are
no narrow photospheric lines and very few strong interstellar absorption
lines.  For the LiF channels, we fit a Gaussian profile to the interstellar
\Ctwo\ 1036.337\AA\ line in each exposure and shifted each exposure to
locate this line at a common velocity. 
For the SiC channels, we 
performed a cross-correlation on the 
the interstellar Lyman lines between 915\AA\ and 930\AA\ (fitting
the \Ctwo\ 1036.337\AA\ line gave essentially identical results).
In order to have enough signal to measure line positions, we discarded
spectra for which the flux was below half of the nominal value for
its channel.

\subsection{GHRS Observations}
Archival \hst/GHRS spectra were used to determine the velocity
structure of the intervening gas and to measure \LA\ absorption
of both \Hone\ and \Done.  A log of the GHRS observations is given in Table
\ref{tab-ghrsobs}.  
All of the data were acquired after the installation of the
Corrective Optics Space Telescope Axial Replacement (COSTAR) unit.
The echelle-mode data, all of which were acquired through the large
science aperture (LSA; 1.74\arcsec $\times$ 1.74\arcsec), 
have a line spread function
with a narrow core of resolution 3.5 \kms\ (FWHM) and broad wings
($\sim$15 \kms\ FWHM) with relatively low power (see Appendix B of 
\citealt{Howk:1999}).

Our reduction of these data follows that of \citet{Howk:1999}.  The
basic calibration makes use of the standard CALHRS pipeline with the
best calibration reference files available as of the end of the GHRS
mission.  The CALHRS processing includes conversion of raw counts to
count rates and corrections for particle radiation contamination, dark
counts, known diode nonuniformities, paired pulse events and scattered
light.  The wavelength calibration was derived from the standard
calibration tables based upon the recorded grating carrousel position.
The absolute wavelength scale should be accurate to $\sim$1 resolution
element, or $\sim$3.5 \kms\ for the echelle-mode data used in this work.

All of the echelle-mode observations employed here were taken at four
separate grating carrousel positions to place the spectrum at
different positions along the 500-diode science array of the GHRS.
This is done to mitigate the effects of the fixed-pattern noise that
can limit the signal-to-noise ratio and produce artifacts in GHRS
spectra.  We have co-added each of the individual sub-exposures by
aligning the interstellar lines of interest (e.g., \Done, \Mgtwo, and 
\Fetwo) in each of the observations.  We used the algorithm of 
\citet{Cardelli:1994}
to explicitly solve for the fixed-pattern noise
spectrum in each of the datasets.  However, we found it unnecessary to
apply this correction because the signal-to-noise ratio of the data
did not necessitate the removal of the low-power features and no
prominent features were found (see discussion in \citealt{Howk:2000}).
We note that the coaddition of the shifted datasets reduces the
strength of any fixed-pattern features by a factor of four.

The inter-order scattered light removal in GHRS echelle-mode data
discussed by \citet{Cardelli:1990, Cardelli:1993} is based upon extensive
pre-flight and in-orbit analysis of GHRS data and is used by the
CALHRS routine.  The coefficients derived by these authors are
appropriate for observations made through the small science aperture
(SSA).  Corrections to these coefficients are often required to bring
the cores of strongly saturated lines observed through the LSA to the
correct zero level.  Typically this is done by adjusting the 
spectrum zero level by an amount equal to a fraction of the average net
flux in each order.  This ``d''
coefficient \citep{Cardelli:1990, Cardelli:1993} accounts for the
residual effects of the local echelle scattering.  

For the Ech-B data covering \Mgtwo\ and \Fetwo, no strongly saturated
lines were available for evaluating the quality of the background
removal, so we adopted the standard d-coefficients from the CALHRS
processing.  The \LA\ absorption profile for \HZ\ should be
flat-bottomed at zero flux near line center.  Although the presence of
strong \LA\ airglow confuses the matter somewhat, it is clear
that the use of the standard d-coefficient (d=0.052) over-subtracts
the core of the \LA\ profile.  We have used the observed flux
level over the wavelength range 1215.74 to 1215.83\AA, assuming this
represents the true zero point, to adjust the d-coefficient.  The
adopted d-coefficient is d = -0.006$\pm$0.033.

The G160M observations of the 1192--1228\AA\ range were acquired through
the Small Science Aperture (0.22\arcsec$\times$0.22\arcsec; SSA).
These data have a resolution of $\sim$19.6\,\kms\ (FWHM).  The scattered
light properties of this holographically--ruled first--order grating
are excellent.  No corrections to the CALHRS background subtractions
were necessary.


\section{Analysis}

\subsection{Method}
\label{sec-method}
We attempted to determine column densities by more than one method 
whenever possible.  The two primary methods were
profile fitting and curve-of-growth analyses.  These two methods 
differ in their sensitivity to some of the sytematic uncertainties
that affect column density measurements.  
A curve-of-growth
analysis was not feasible for \Hone, as essentially all of the lines
other than \LA\ are on the flat part of the curve of growth.
Instead, \Hone\ column density measurements based on fits to the EUV spectrum
of \HZ\ were taken from the literature to provide a consistency check on
our results.

Profile fitting was used to determine the column densities of \Hone,
\Done, \Oone, and \None.
The profile fitting was performed using the program Owens.f, written
by one of us (M.L.).  This program uses \chisq\ minimization to adjust
temperatures, turbulent velocities, and relative
velocities for multiple gas components, to adjust column densities for
each species present in each gas component, and to adjust polynomial fits
to continuum fluxes, line spread functions, background levels, etc.
for each specified spectral region.  
Each quantity can be adjusted by the program
or held fixed to user-specified values.  Further information on this
program is provided in the companion paper by \citet{Hebrard:2001}.

The \fuse\ and \ghrs\ spectra were normalized prior to fitting by means of
an NLTE stellar model.  
The continuum in the vicinity
of individual absorption lines was then fit by a low-order polynomial as
required to account for any residual flux calibration errors or flat-field
effects.  
We use the program TLUSTY \citep{Hubeny:1995}
to calculate the atmospheric structure of \HZ\ under the
assumption of NLTE, and then we compute the detailed emergent flux
using the program SYNSPEC (Hubeny 2000, private communication).
The hydrogen line profiles are obtained from the Stark
broadening tables computed by \citet{Lemke:1997} using the
program of \citet{Vidal:1973}.  We use a NLTE model
atmosphere to calculate the emergent flux, because NLTE effects are
important in the core of the Ly$\,\alpha$ line, which is formed high
in the atmosphere where departure from LTE is significant (see, e.g.,
\citealt{Wesemael:1980}).
The model parameters used were as discussed in Section \ref{sec_los}:
the composition was pure hydrogen, \Teff = 50900K, and \logg = 8.00.
The model spectrum was then
normalized to V = 12.914 \citep{Bohlin:1995}.  
Except for NLTE effects in 
the narrow cores of the Lyman lines it is essentially identical to
the LTE model described in detail by
\citet{Kruk:1999} and references therein.  

The \LA\ profile was also fit using a different program written by one of us 
(J.C.H), and the profile fitting code of Ed Fitzpatrick 
\citep{Spitzer:1993} was used
to provide a consistency check for certain test cases. 

We have derived column densities for \ion{N}{1} and \ion{O}{1} by
fitting the measured equivalent widths with a single-component
Gaussian (Maxwellian) curve of growth \citep{Spitzer:1978, Jenkins:1986}.
The analysis was performed using software adapted from that
described by \citet{Savage:1990}, which varies the Doppler
parameter, $b$, and column density so as to minimize $\chi^2$ between
the observed equivalent widths and a model curve of growth.
The stellar continuum in the vicinity of each
line was estimated using a low-order Legendre polynomial fit to the
data.  The dispersion of the data about the fit was used to estimate
the uncertainties in the final integrated equivalent width
measurements as discussed by \citet{Sembach:1992}.
This estimate includes contributions from both Poisson noise and
uncorrected high-frequency fixed-pattern detector features.  The
measurement uncertainties quoted include contributions from this noise
estimate, uncertainties in the Legendre fit parameters 
\citep{Sembach:1992},
and estimates of the systematic uncertainties inherent
in the continuum placement and velocity integration range.
Atomic data for both the profile fitting and the curve of growth analysis
were taken from \citet[Morton 2001, private communication]{Morton:1991}.


The integrated equivalent widths for several interstellar species are
presented in Table \ref{tab-eqwidths}, and the results from the 
curve-of-growth fits are given in Table \ref{tab-cogcolumns}.
For most species other than \ion{N}{1} and \ion{O}{1} the number of
available transitions was insufficient to fit a curve of growth and
determine a b-value; instead, the measured equivalent width was combined
with a curve of growth computed using the \ion{O}{1} b-value to derive
the column density.  For these cases the table also lists
the difference between the column
density derived in this manner and that derived using a straight integration
of the apparent column density profiles \citep{Savage:1991},
providing a measure of the saturation correction.

\subsection{Systematic Uncertainties}
\label{sec_syst}
Column density determinations are subject to numerous possible systematic
uncertainties.  These are described in the following section,
along with brief descriptions of how these were addressed.  The
magnitude of each effect on the column densities will then be addressed
in the discussion of each species as appropriate.

\subsubsection{Overlapping velocity components along the line of sight}
One of the major difficulties typically encountered in measuring column
densities is accounting for overlapping absorption profiles resulting
from clouds with similar velocities.  As we will show below,
there appears to be only a single absorption component along the line
of sight to \HZ, so this complicating factor is not present.

\subsubsection{Narrow absorption features in the stellar spectrum}
Stellar photospheres ordinarily contain trace quantities of numerous
elements that can give rise to narrow absorption features in the
emergent spectrum.  In some cases the stellar spectrum is well-understood
and the presence of such narrow features can be predicted and possibly
even modelled.  In other cases, the wavelengths of the various transitions
are poorly known and it is impossible to be certain how much of an
absorption profile is due to interstellar gas and how much results from the
photospheric spectrum.  However, as discussed in Section \ref{sec_los},
the atmosphere of \HZ\ appears to be composed solely of hydrogen.
Thus, this potential complicating factor is also not a concern for 
our analysis of \HZ.

\subsubsection{Uncertainties in the stellar \LA\ profile}
The high gravity of \HZ\ causes the photospheric Lyman lines to be quite
broad, hence the stellar spectrum is varying slowly in the vicinity of
all of the interstellar lines being fit other than \Hone\ \LA.  Because we
are fitting the damping wings of the interstellar \Hone\ \LA\ profile,
spanning roughly 6\AA, we need to be able to model the photospheric
absorption very accurately.  In addition, the \Hone\ column  density
along the line of sight to \HZ\ is sufficiently low, and the relative
velocity of the star and interstellar gas is sufficiently high,
that the NLTE core of the photospheric \LA\ profile may affect the red wing
of the combined profile.
\HZ\ has been studied in great detail, however,
and model atmospheres and synthetic spectra of \HZ, along with those
of several similarly well-studied DA white dwarfs, are used to define
the flux calibration for most orbiting UV spectrometers
\citep[and references therein]{Bohlin:2000}.  The parameters describing
\HZ\ are discussed in Section \ref{sec_los}, and the fits to the \LA\ profile
are discussed in Section \ref{sec-h_one}.

\subsubsection{Continuum placement errors}
Errors in placement of the continuum or the background will clearly
affect the column density determination.  In most regions of the
spectrum the continuum appears to be smooth and slowly-varying in
the vicinity of each of the lines measured.  If the
continuum did not appear to be well-behaved, most likely because of
large-scale fixed-pattern noise effects in the detector, then the
the affected line was not included in the fit (the SiC1B MDRS measurement
of the \Oone\,948.686\AA\ line is such a case).  Uncertainties in
continuum placement are included explicitly in the equivalent width
measurements used in the COG analysis, and implicitly in the \chisq\ 
minimization performed by the profile fitting.  This latter approach is adequate
if the true continuum is in fact well-represented by a low-order
polynomial, and if the absorption being measured occupies a small fraction
of the region being fit.  For the broad \LA\ profile a
variety of continuum fits were tested to determine the sensitivity
of the results to the adopted continuum.  This will be discussed in
greater detail below in the section on the measurement of the \Hone\ column
density.

\subsubsection{Background placement errors}
The dominant background for \fuse\ spectra of a bright source 
such as \HZ\ is grating
scattering of source flux.  The intrinsic detector background and
diffuse stray light in the spectrograph are negligible in comparison.
The grating-scattered source flux has been characterized by measuring
the flux in the cores of highly-saturated \Hmol\ absorption line
profiles in the spectra of stars with translucent clouds along the line
of sight.  Typical results for this background are that the
level is about 1\% of the continuum flux in the SiC channels,
and 0.3\% to 0.5\% of the continuum flux in the LiF channels.
In the \HZ\ \fuse\ observations, the flux shortward of the Lyman limit
is indeed about 1\% of the continuum flux in the SiC1B channel.
None of the interstellar absorption lines in 
the \HZ\ spectrum are broad enough for
the true background level to be seen in the line cores.  The instrumental
LSF has a broad component that has a FWHM of about
18 pixels, which is broad enough that even the \Hone\ Lyman line cores
do not reach the level of the background.  
This point is discussed further in Section \ref{sec_lsf}.  The effects
of uncertainty in the background level were investigated by performing
the profile fits with the background level set to zero, set to our
best estimate of the background, and and set to twice our best
estimate of the background.  The resulting column density variation
was typically 0.01 dex, and is included in the quoted uncertainties
for the column densities.  The background effects in the GHRS spectra
are discussed in the section below on the \Hone\ column density
measurement.

\subsubsection{Geocoronal emission}
\label{sec_airglow}
Contamination of the spectrum by geocoronal emission can be significant for
the large spectrograph apertures employed by \fuse.  Such emission is quite 
strong at \LB, but declines rapidly in strength for the higher Lyman lines.
Geocoronal emission is also present on the day side of the orbit for many
of the same \Oone\ and \None\ transitions for which we are trying to measure
interstellar absorption.  The widths of geocoronal lines are limited
by the size of the spectrograph apertures.  For the LWRS aperture this width
is typically 0.35\AA, or the equivalent of 106\,\kms\ at \LB, while for the MDRS
aperture this width is about 0.045\,\AA, or the equivalent of 13\,\kms\ at \LB.


The MDRS aperture is sufficiently narrow ($\sim$13\,\kms) that the 
geocentric velocity of the
interstellar gas shifts the geocoronal emission almost completely away from
narrow lines.  The red wing of the broad interstellar \LB\ line is filled
in by geocoronal emission, but no difference is seen in the \Oone\ or \None\ 
absorption profiles when the day and night data are analyzed separately.
The absorption profiles measured in the LiF1  channel are actually deeper
for the day data than the night data, presumably because the LiF1
grating is more stable in the day than in the night and the effective 
resolution is correspondingly better.

\subsubsection{Uncorrected fixed-pattern noise}
Uncorrected fixed-pattern noise on small scales can lead to significant errors
in the shapes and fluxes of line profiles; 
on larger scales it can lead to continuum placement
errors.  The \fuse\ detectors exhibit significant fixed-pattern noise on
both small and large scales, caused primarily by the inherent granularity
of the microchannel plates (MCPs).  The slowly-varying relative spacing of 
the pores in the stack of 3 MCPs gives rise to Moire fringe effects with
a spacing of 8-9 pixels, comparable to the width of the LSF.  Changes in
pore spacing at fiber bundles and regions of dead pores can cause 
changes in the detector quantum efficiency on larger scales.
The present CALFUSE pipeline flags pixels affected by dead regions of the
detector, but otherwise fixed-pattern features are not yet corrected.

We have mitigated the effects of fixed-pattern noise by several means.
Every transition being examined in the \fuse\ bandpass is sampled by either
2 or 4 channels, so we have verified that each such measurement provided
essentially similar results.  In addition, because \HZ\ was observed in both
LWRS and MDRS apertures, we have checked for consistency of the 
results.
The MDRS data were obtained with an FP-split procedure, which substantially
reduces the fixed-pattern noise on scales of one to two resolution elements.
The resulting spectra typically have a S/N that is about 80-90\% of that
expected from photon statistics, depending on the size of the region examined,
primarily because the larger-scale fixed-pattern noise has not been
removed.  However, most of these fluctuations are slowly varying and
can be removed by low-order polynomial fits to the continuum in the vicinity
of absorption lines of interest.

\subsubsection{Uncertainties in the instrumental line spread function}
\label{sec_lsf}
Characterization of the \fuse\ LSF is still
in a preliminary state.  There appears to be both a narrow and a broad
component to the LSF.  The narrow component is somewhat wavelength-dependent
and is sensitive to the accuracy of the correction for grating motion
during exposures and to the accuracy with which the user was able to
co-align the spectra from each exposure prior to combining them.  
In addition, data acquired in histogram mode cannot be corrected for the
degradation of the spectral resolution caused by rotation of the 
primary mirrors or gratings during an exposure.
For example, in some channels the resolution is better during the night
portion of the orbit and for other channels it may be better during
the day.  As a result,
the effective width of the narrow component of the LSF will vary from
one observation to another.  We measured the effective width of
this narrow component by fixing the temperature and turbulent velocity
of the interstellar gas to that found from the GHRS measurements (see below),
and letting the LSF widths vary in the profile fits.  The results
were generally consistent from one absorption line to another, and
were typically about 9 pixels FWHM (corresponding typically to 0.062\AA\ 
in the LiF channels and 0.057\AA\ in the SiC channels).  
This is consistent with what is
found from other FUSE datasets.  These widths were then held fixed when
performing subsequent fits.  It should be noted that the LSF width is
significantly greater near the edges of the detector active area, as
are the uncertainties in the wavelength calibration.

The broad component of the LSF appears to be somewhat less sensitive to 
the effects of
grating rotation, primary mirror rotation, or coalignment of spectra.
The initial characterization the available data indicates that about
30 percent of the area of the LSF is in a component with a width in the
range of 17-24 pixels.  Analysis of highly saturated lines that are 
flat-bottomed over regions of 150 pixels or more reveals that there is no
significant component of the LSF with a characteristic width larger than
25 pixels.
The effects of this broad component of the LSF on the profiles of unsaturated
lines may not be evident unless the signal-to-noise ratio of the data
is very high; however, its main effect is to introduce a substantial
apparent increase in the background under the line.  If a weak line is fit
with a single, narrow, Gaussian LSF, the background level in the fit
must be increased to account for this effect.  
In the profile-fitting analysis of the \HZ\ ISM absorption we have 
represented the LSF
as a two-component Gaussian function.  The width of the narrow component
was determined to be about 9 pixels as described in the previous paragraph,
and the broad component width was set to 17 pixels, with an amplitude
corresponding to 30\% of the total area.

If the LSF has very long shallow tails that cannot be distinguished from
the continuum, then both the profile fitting technique and curve-of-growth
analysis would likely be subject to continuum placement errors that will cause
some of the area of the line to be missed and therefore cause the 
the column density to be underestimated.  However, while there is a component
of the LSF that has roughly twice the width of the narrow component, there
is no component with very long extended tails.
In the absence of such a weak and very broad component, 
the curve-of-growth analysis is independent of the shape of the LSF, and thus
provides a valuable check that the profile fit was not adversely affected
by errors in the assumed LSF.

The LSF of the GHRS modes used here is smaller than the Doppler widths of
the lines being measured, other than for \Fetwo,
so there is no significant systematic error associated with uncertainties 
in this LSF.
The Doppler width of the \Fetwo\ 2382.765\AA\ line is comparable to the LSF
width, so the determination of the Doppler width for this line
does have relatively large  uncertainties.
The column density, however, is not adversely affected because the line is
so weak.

\subsubsection{Uncertainties in the f-values}
Errors in the atomic data used in the fits have the potential of biasing the
results.  If such errors are distributed randomly then the use of numerous
lines will minimize any systematic bias of the overall column density
determination.  If the oscillator strengths for a given species are all 
too large or too small then the estimate for the column density will
be biased.  However, such a systematic error in the f-values seems unlikely
to result in a consistent fit to all of the lines, especially
if lines are included that are not on the linear part of the curve of
growth.  We have tested for effects on the derived column density
resulting from errors in the f-values by repeating the
profile fits after removing each transition and verifying that the 
derived column density did not vary by more than 1\,$\sigma$.
An examination of the curve-of-growth plots also shows that a satisfactory
fit can be obtained that is consistent with each of the included points.
The only transition that was found to have an f-value clearly inconsistent
with those of the other transitions was the that of the \Oone\ 1026.473\AA\ line
(see \citealt{Sembach:2000}).
This line was excluded from all fits and from the curve-of-growth analysis.

\subsection{Results}

\subsubsection{Line of sight velocity structure}
\label{sec-losvel}
\begin{figure}
\plotone{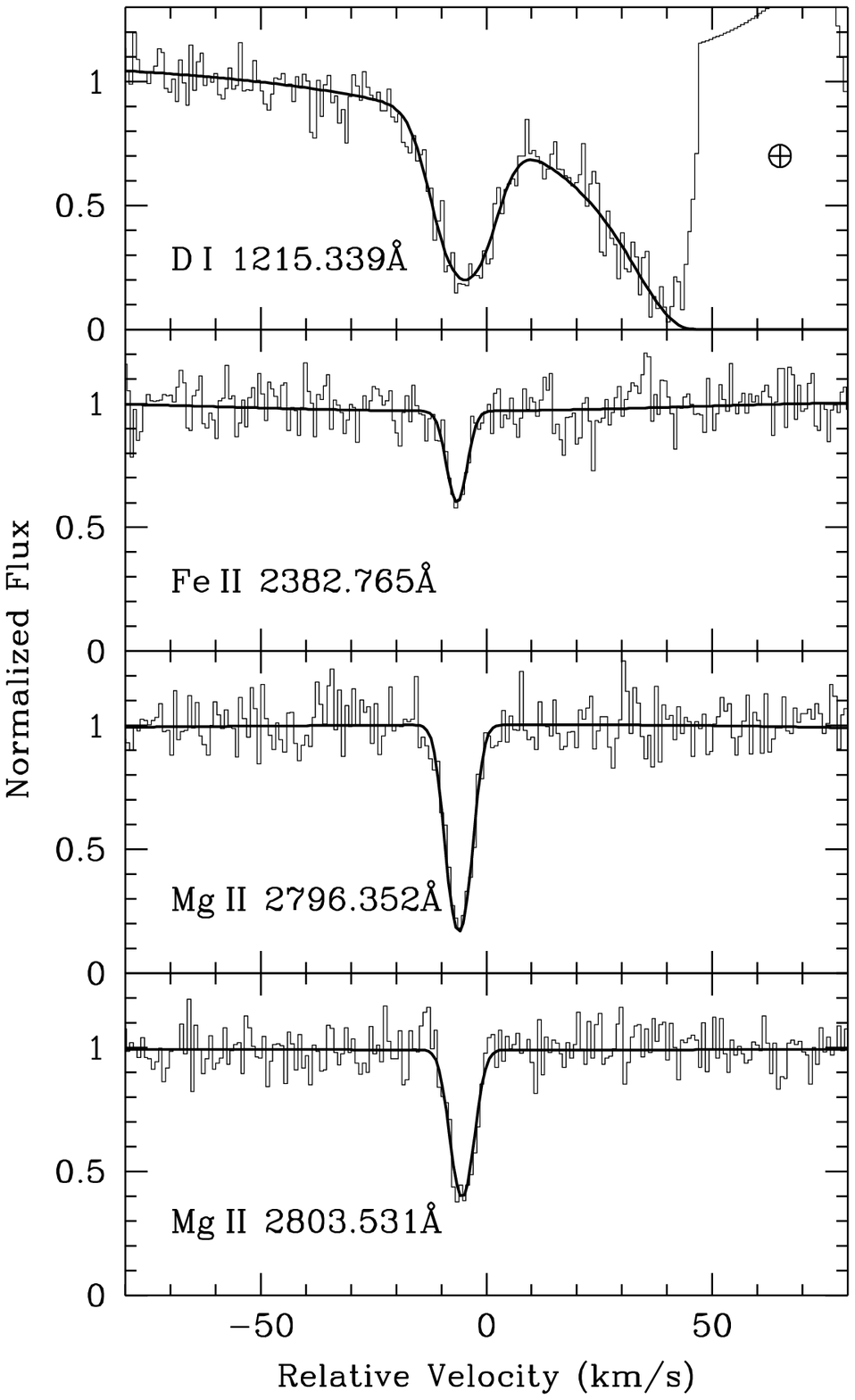}
\caption{Absorption profiles of \Done, \Fetwo, and \Mgtwo\ are
plotted as a function of heliocentric velocity.  The histograms are the
GHRS echelle measurements, and the smooth line is the single-component fit.
No evidence for a second velocity component is seen.  The \Done\ line
sits on the broad wing of the interstellar \Hone\ \LA\ profile; the
truncated peak at +65\,\kms\ in the \Done\ plot is geocoronal \LA\ emission.
\label{fig_ghrs_v}   }
\end{figure}

The velocity structure of absorbing gas along the line of sight was
determined from GHRS observations of \Done\ 1215.339\AA,
\Fetwo\ 2383.765\AA, \Mgtwo\ 2796.352\AA, and \Mgtwo\ 2803.531\AA\ line
profiles.  These lines are well-resolved, and are isolated from any
other absorption features.  We obtained a satisfactory fit to the
lines with a single absorption component with a heliocentric velocity
of -5.2~$\pm$~0.1\kms, a temperature of T~=~5353~$\pm$~948~K, and a
non-thermal velocity of \vnt~=~2.0~$\pm$~0.2\kms\ (the uncertainties
are 1\,$\sigma$ statistical uncertainties; the systematic uncertainty
on the absolute velocity is $\pm$ 3\,\kms).  The total Doppler
parameter, $b$, is related to the temperature and non--thermal
velocity parameter such that $b^{2} = 0.0165\,T/A~+~\vnt^{2}$.  The
observed profiles and the best model fits are shown in Figure
\ref{fig_ghrs_v}.  There is no evidence for an additional velocity
component.  These values for T and \vnt\ were not held fixed in
subsequent fits (e.g. for \Oone\ or \None).  Instead, the derived
Doppler widths for each species were computed and compared with the
ranges allowed by the above values; in each case the derived Doppler
widths were found to be consistent with the 1\,$\sigma$ uncertainties
quoted above.  The b-values for \Oone\ and \None\ predicted from the
values of T and \vnt\ given above are: b(\Oone) = 3.1 $\pm$ 0.2\,\kms
and b(\None) = 3.2 $\pm$ 0.2\,\kms\ (1\,$\sigma$ uncertainties).

\subsubsection{\Hone\ Column Density}
\label{sec-h_one}
The interstellar \Hone\ column density along the line of sight to \HZ\ can
be determined by three different means:  fitting the spectral shape of
the EUV continuum, fitting the damping wings of the \Hone\ \LA\ profile,
and fitting the converging Lyman series.  This last method has the
greatest uncertainties because the lines are typically all saturated, but
for the low column density of this line of sight it is possible to obtain
meaningful results.

Determination of the \Hone\ column density by analysis of the EUV
spectrum of a star requires accurate knowledge of the stellar continuum
flux.  The EUV spectrum of a star is very sensitive to the presence of
even trace quantitites of heavy elements in the photosphere, and substantial
effort is usually required in order to model this spectrum accurately.  \HZ\ is
unusual in that no traces of heavy elements have yet been detected in
its photosphere; hence it is not affected by this potential source of
systematic error.  \euve\ spectra of \HZ\ have been analyzed by several
groups, using different observations, different model atmosphere codes,
and different fitting methods.  The three most recent analyses of \euve\ 
observations will be described briefly below.

\citet{Dupuis:1995} analyzed observations
from 1994 March 25,28 and 1994 May 20.  They first determined the \Heone\
and \Hetwo\ column densities by fitting the ionization edges at 504\AA\ 
and 228\AA, respectively, and then determined the stellar \Teff\ and
interstellar \Hone\ column density by fits over the 400--550\AA\ bandpass.
This wavelength range was chosen because it was least sensitive to variations
in surface gravity or heavy element abundance.  The fitting was done with
model spectra computed from pure H model atmospheres \citep{Vennes:1992b}.  
The results were \Teff\ = 51100\,K $\pm$ 500\,K, and 
\Nh\ = $(8.7 \pm 0.6)\times 10^{17}$\,\cmsq\ (\onesig).
They also estimated that systematic uncertainties in the effective area of
20\% would change \Teff\ by 3\% and \Nh\ by 10\%.  Uncertainties in the
V magnitude (used to normalize the models) would have similar effects, but
are small in comparison.  The ionization fraction of hydrogen was estimated
by assuming that the ratio of total H to total He was 10:1, that there
was no appreciable \Hethree, and comparing the measured \Nh\ to the
measured \Heone\ and \Hetwo\ column densities.  The resulting ionization
fraction for H, $f_H$ = N(\Htwo) / N(H), was $<$ 0.4 $\pm$0.1.

\citet{Barstow:1997} analyzed the 1994 May 20
observation also, but used H+He models computed by \citet{Koester:1991}
and determined all stellar and ISM parameters from a simultaneous fit
to the entire \euve\ spectrum.  The values for \Teff\ and \logg\ were
constrained to remain within the \onesig\ uncertainties of the values
determined by \citet{Napiwotzki:1993}.  The V magnitude used, 12.99,
corresponds to model fluxes that are fainter by a factor of 1.077 than
would be the case if the present best value of V = 12.909 were used
\citep{Bohlin:2000}.  This would result in an underestimate of
the derived \Nh\ column density by a few percent.  They performed fits
with two classes of models:  homogeneous H+He mixtures, and stratified
compositions.  The derived interstellar \Hone\ column densities 
and \onesig\ uncertainties were
\Nh\ = $(8.8 \pm 0.2) \times 10^{17}$\cmsq\ (homogeneous) and 
\Nh\ = $8.3 \err{0.2}{0.1} \times 10^{17}$\cmsq\ (stratified).
The difference between the results for the two models exceeds the
statistical errors in the fits, indicating that 
systematic uncertainties in the nature of the appropriate model are
significant.  The ionization fraction was derived in the same manner as
by \citet{Dupuis:1995} to be $f_H$ = 0.19\err{0.15}{0.19} for homogeneous
models, and $f_H$ = 0.22\err{0.11}{0.14} for stratified models.
We have assumed that the results presented by \citet{Barstow:1997}
supercede the earlier results in \citet{Barstow:1995}, so we will not
discuss the earlier results here.

\citet{Wolff:1999} analyzed observations obtained
1997 June 25,26.  They first fit the \Heone\ and \Hetwo\ ionization edges
to determine the corresponding column densities, and then fit the
full spectrum to determine the \Hone\ column density.  The stellar spectrum
was computed using the LTE model atmosphere code of \citet{Koester:1996},
assuming a pure hydrogen composition.  The atmosphere parameters were
not free, but fixed at values derived from fits to the Balmer lines.
The values used for \Teff\ were 50800 $\pm$ 300\,K, and the corresponding 
column densities were \Nh\ = 8.851\err{0.311}{0.300} $\times 10^{17}$\,\cmsq.
The uncertainties in the column densities are dominated by the uncertainty
in the effective temperature (Wolff 2001, private communication).  
The V magnitude used to normalize the models
was 12.89; this would overestimate the stellar flux by about 2\% and
would result in only a very slight overestimate of the \Hone\ column
density.  The ionization fraction was derived in the same manner as
the previous two analyses to be $f_H$ = 0.03\err{0.10}{0.03}.

An EUV spectrum of this star was also obtained by HUT
and analyzed by \citet{Kimble:1993} to derive 
\Nh\ = $(6.5 \pm 0.5) \times 10^{17}$\cmsq\ 
($1\,\sigma$ statistical errors; systematic uncertainties were estimated
at 15--20\% and are dominated by calibration uncertainties).  
This falls below the more recent \euve\ determinations
by approximately the combined random and systematic uncertainties.
These uncertainties are considerably larger than those of the much higher
signal-to-noise \euve\ spectra,  so we will not consider this
measurement further.  The three \euve\ measurements described above will
be addressed again at the end of this section. 

\begin{figure}
\plotone{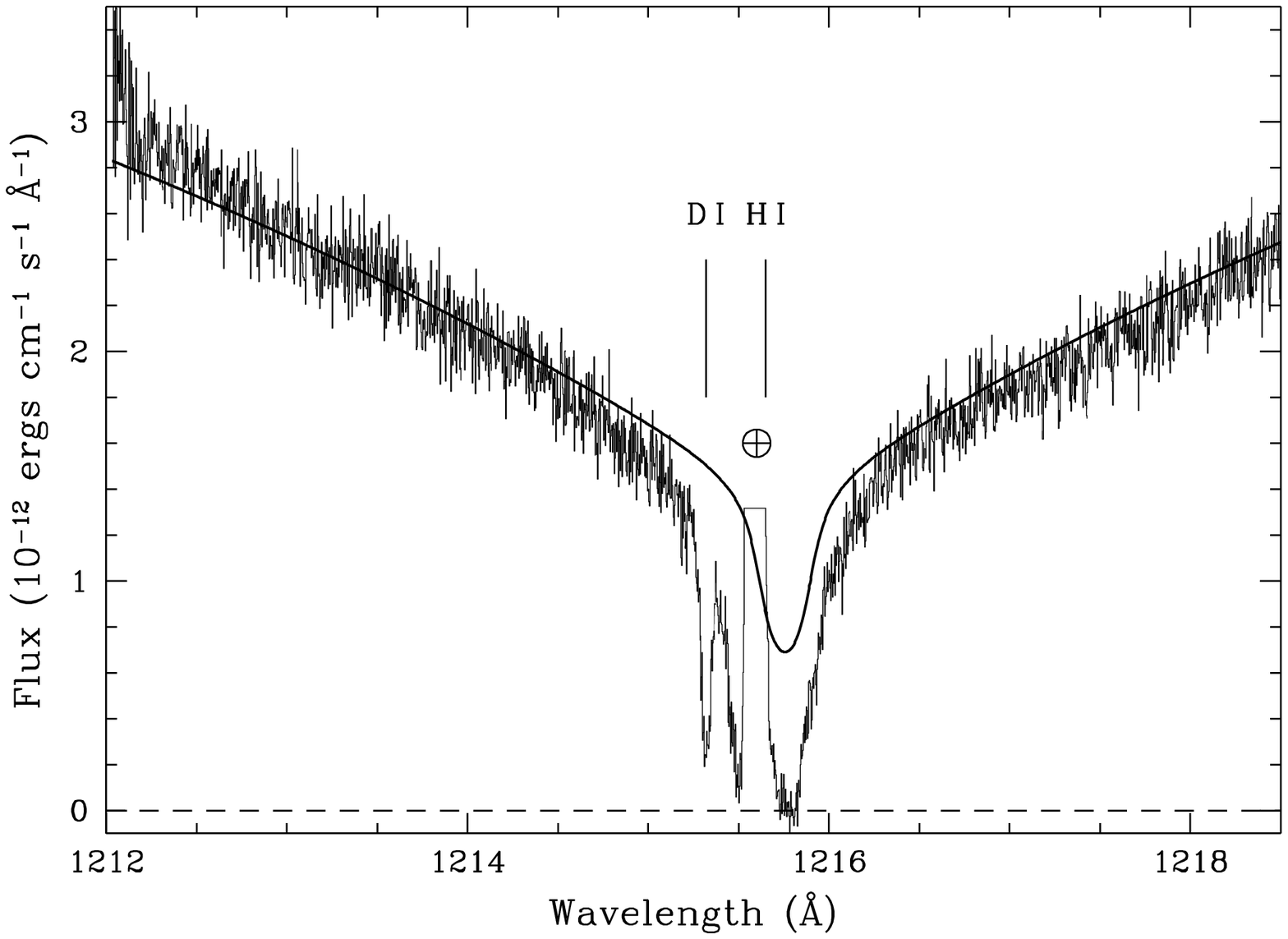}
\caption{The GHRS echelle spectrum of the \LA\ profile 
is shown as the histogram, the stellar model as the smooth curve.  
The model is shifted to a heliocentric velocity of +21\,\kms.  The positions
of the interstellar \Hone\ and \Done\ absorption lines are marked.
The truncated peak at 1215.6\AA\ marked with an earth symbol is geocoronal
\Hone\ \LA\ emission.
The GHRS spectrum was normalized to match the mean flux of the model over
the regions 1212.25\AA-1213.25\AA\ and 1218.0\AA-1218.5\AA; additional
low-order polynomial corrections were applied to the continuum flux
when fitting (see text).
The zero level is shown as the dashed line.
\label{fig_lya_raw}
  }
\end{figure}

We fit the GHRS spectrum of the interstellar \Hone\ plus \Done\ \LA\ profile 
towards \HZ\ 
to determine the properties of the neutral hydrogen in this direction.
The spectrum was first scaled by a factor of 0.866 in order to match our
model flux over the regions 1212.25 -- 1213.25\AA\ and 1218.0 -- 1218.5\AA.
The spectrum was then normalized by an NLTE stellar model prior to fitting (see
Section \ref{sec-method} for a description of the model).
We found that additional normalization by a low-order polynomial was
required to account for potential small wavelength--dependent
errors in the flux calibration.
The apparent photospheric radial velocity of \HZ\ (\vphot) was measured by 
\citet{Reid:1996} to be +21\,\kms.  
The spectrum and the model (shifted by +21\,\kms) are shown
plotted in Figure \ref{fig_lya_raw}.
A visual inspection of the figure shows that at least a linear polynomial
will be required as an additional correction to the flux calibration.

\begin{figure}
\plotone{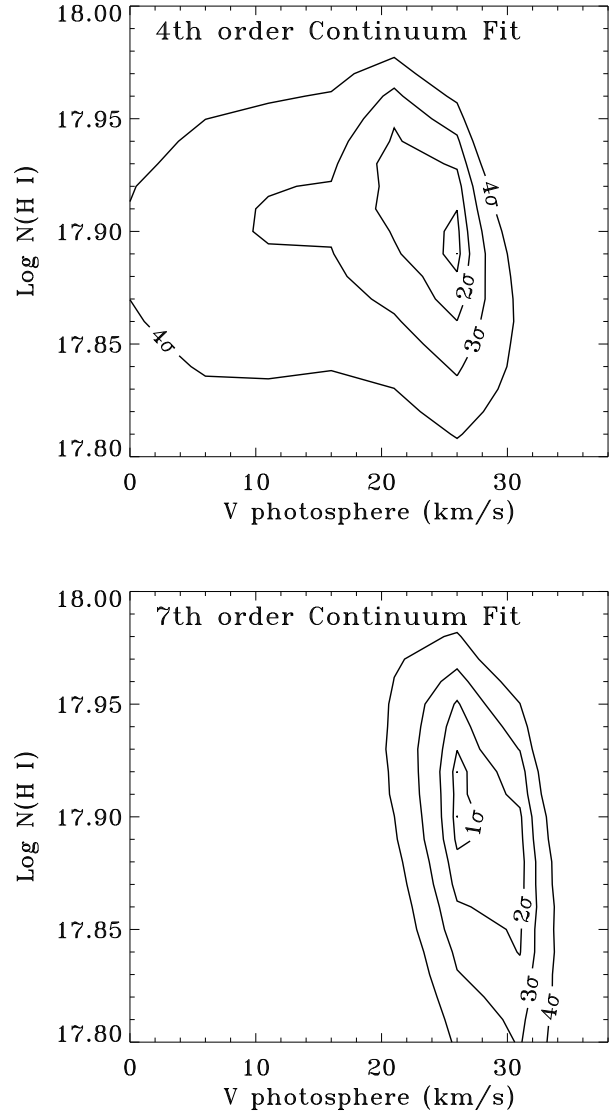}
\caption{Plots of the \chisq\ contours are shown for fits to the GHRS
\LA\ profile as a function of \lognh\ and photospheric velocity.
The top plot shows the results for $4^{th}$ order polynomial continuum fits,
and the bottom plot shows the results for $7^{th}$ order polynomial continuum
fits.
\label{fig_lyachisq}
  }
\end{figure}

No explicit uncertainty was given by \citet{Reid:1996}
for the radial velocity, but it is probably about 10-15\,\kms.
We therefore examined the effects on the derived \Hone\ column density
of varying this velocity.  We created stellar models with the photospheric
velocity set at zero, at the nominal value of +21\kms, and at $\pm$5, $\pm$10,
$\pm$15\,\kms\ with respect to the nominal value.  The GHRS \LA\ spectrum
was normalized by each model in turn, and the best fit to the spectrum
was generated for fixed values of \lognh\ ranging from 17.80 to 18.00
in steps of 0.01 dex.  The resulting \chisq\ contours are plotted in
Figure \ref{fig_lyachisq}.  The \Fetwo\ and \Mgtwo\ lines were fit
simultaneously with the \LA\ \Hone\ and \Done\ line profiles.
Because the \LA\ profile extends over several Angstroms, it may be affected
more by uncertainties in the flux calibration than are narrow lines.
To investigate this possibility,
we generated these \chisq\ contours for continuum fits performed with
both fourth-order and seventh-order polynomials.  
There was little correlation in either case between \lognh\ and the
photospheric velocity.
For the fourth-order fits, the minimum \chisq\ was obtained for a
photospheric velocity of 26\,\kms, but the \chisq\ increased slowly for
smaller velocities and increased rather steeply for higher velocities.
For the seventh-order continuum fits, the minimum \chisq\ was also at 26\,\kms,
but it increased rather rapidly for both lower and higher velocities.
The \Hone\ column density and \twosig\ uncertainty derived was
\lognh\ = 17.895\err{0.045}{0.035} from the fourth-order continuum fits, and 
\lognh\ = 17.905\err{0.045}{0.065} from the seventh-order fits.
The corresponding \chisq\ values were 2750 for 2871 degrees of freedom,
and 2711 for 2868 degrees of freedom, respectively.
Based on these fits, we have adopted the value \lognh\ = 17.90\err{0.05}{0.06}
for our combined GHRS and \fuse\ datasets.

We examined the behavior of the continuum fits by varying
the order of the polynomial from 1 to 8,  for
photospheric velocities ($\vphot$) of 0, +5, +11, +21, and
+31\,\kms, and allowing the \Hone\ column density to vary freely in each fit.
For velocities of 0 and +5 \kms\ there
was relatively little effect of polynomial order on the reduced \chisq\ (for
orders $>$ 1).  For higher velocities the \chisq\ improved significantly
for orders up to 4 and improved only slightly thereafter (with minima
at order 7), except that
for +31\,\kms\ the \chisq\ dropped again for order 8.  
An examination of the polynomials in each case
show that the dominant term gradually changes from linear at low \vphot\ to
cubic at high \vphot, 
and that the amplitudes increase from $\pm$ 2\% at low \vphot\ to $\pm$ 3\%
at higher \vphot.  The higher-order terms gave slight improvements to
the overall \chisq, but did not change the character of the continuum fits.

We evaluated the effects of the uncertainties of the background
flux level on the derived \lognh.  
Our best estimate of the background level at \LA\ was
0.5\% of the continuum flux level, obtained by letting the background be
a free parameter in the fit.  We then varied the background level by +2\% and
-2\% of the continuum flux level, causing \lognh\ to change by 
only $\pm$ 0.012.

We were unable to obtain a satisfactory fit to the observed \LA\ profile
with a single component of neutral material, but found that a good fit
was possible only if a second, low-column density absorber were included.  
The temperature, column density, and relative velocity of this component 
were free parameters in the fits to the \LA\ profile used to construct the
\chisq\ contours described above.  
For the fits in which the continuum 
was represented by fourth-order polynomials, the column density of this
second component was \lognh\ = 14.9\err{0.3}{0.5} (\twosig).
For the fits in which the continuum 
was represented by seventh-order polynomials, the column density of this
second component was \lognh\ = 14.9\err{0.6}{0.2} (\twosig);
We will adopt the value \lognh\ = 14.9\err{0.6}{0.5} (\twosig).
The temperature of
this component was 30,000\,K $\pm$ 10,000\,K (\twosig) for both cases.
As the column density of the main \Hone\ component was increased,
the derived column density for this second component would also increase,
and its temperature would decrease.  Similarly, as \vphot\ was increased,
shifting the NLTE core of the photospheric Lyman line closer to the
red side of the saturated portion of the interstellar absorption profile,
the column density of this second component would increase and the
temperature would decrease.
The heliocentric velocity of
this component was found to be about -2.0\,\kms, but as $\vphot$ was
varied from it would shift in the opposite direction (i.e. it would
increase as $\vphot$ decreased).
We suggest that this low-column density, high-temperature \Hone\ component
may be due to the
``hydrogen wall'' about the Solar System.
The hydrogen wall is an accumulation of gas at the boundary of the
heliosphere, where the solar wind and LISM interact.  
Hydrogen walls have been detected around the Sun along several lines
of sight, and around several other stars 
\citep{Wood:1998,Linsky:1996, Wood:2000}.
The gas in these walls is fairly hot, 30,000K - 180,000K, and \Hone\ column
densities range from $\lognh = 13.75$ to $\lognh = 14.75$.
The properties of the second \Hone\ component in our fits are consistent
with these other detections.
One potential problem with this interpretation, however, is
that the \LA\ profile observed for
the similar high-latitude sightline to 31\,Com ($l = 115\degr, b = +89\degr$)
can be fit very well without any high-temperature component \citep{Wood:2000}.
The models of the interaction of the heliosphere and the LIC described
by \citet{Wood:2000} do predict absorption by such a hot component
for the line of sight to 31\,Com, so one possible explanation is that the
LIC is inhomogeneous on small scales and that the corresponding conditions
at the boundary of the heliosphere vary accordingly.

\begin{figure}
\epsscale{0.8}
\plotone{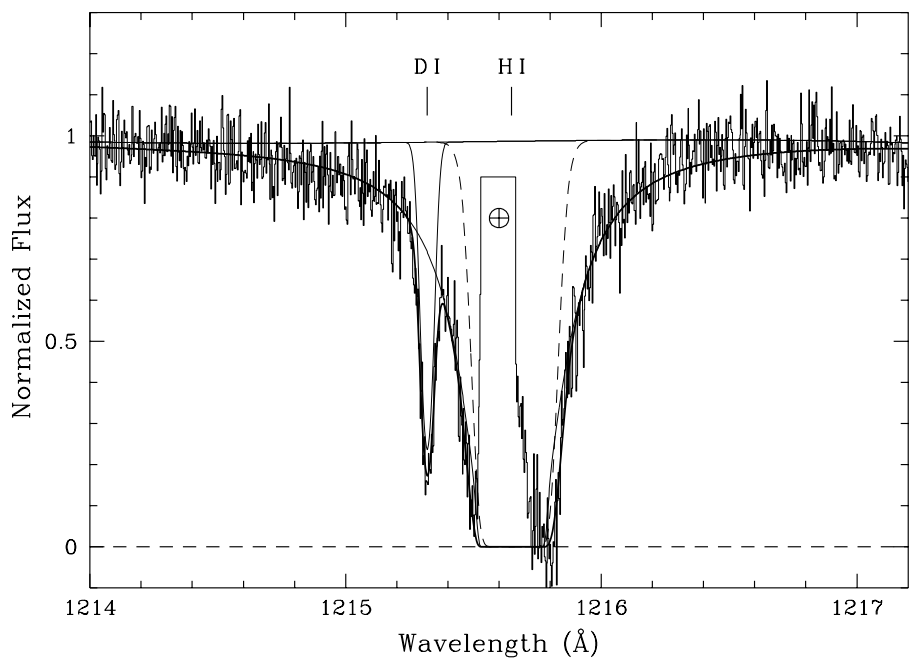}
\caption{The normalized GHRS spectrum of the \LA\ profile 
is shown as the histogram, and the best fit is shown as the smooth
curve.  The thin curves show the continuum fit and the individual
\Hone\ and
\Done\ profiles; the dashed curve shows the low-column density high-temperature
\Hone\ component; and the thick solid line shows the overall fit.
The stellar model in this fit was shifted to a heliocentric velocity of
26\,\kms, which provided slightly better fits than the nominal value of
21\,\kms.
The positions
of the interstellar \Hone\ and \Done\ absorption lines are marked.
The truncated peak at 1215.6\AA\ marked with an earth symbol is geocoronal
\Hone\ \LA\ emission.
The zero level is shown as a dashed line.
\label{fig_lyanorm}
  }
\end{figure}

The best fit to the normalized \LA\ profile, including the main component
and the high-temperature component, is shown in Figure \ref{fig_lyanorm}.
The relative velocity of the main absorbing component matched that found
for the \Done, \Mgtwo, and \Fetwo\ lines, about -5\,\kms.  This 
differs significantly from that expected for the 
local interstellar cloud (LIC), \mbox{-8.8\kms,}
and the column density is much higher than would be expected
from the LIC along this line of sight ($\lognh = 16.47$; 
\citet{Redfield:2000}).  The velocity differs even more from that
expected for the `G' cloud, -12.1\,\kms.
The velocity is consistent, however, with that found for the single absorber
along the line of sight to 31\,Com by \citet{Piskunov:1997} (-4 $\pm$ 1\,\kms).
This absorber is called the North Galactic Pole cloud by 
\citet[see their Figure 1]{Linsky:2000}.

\begin{figure}
\plotone{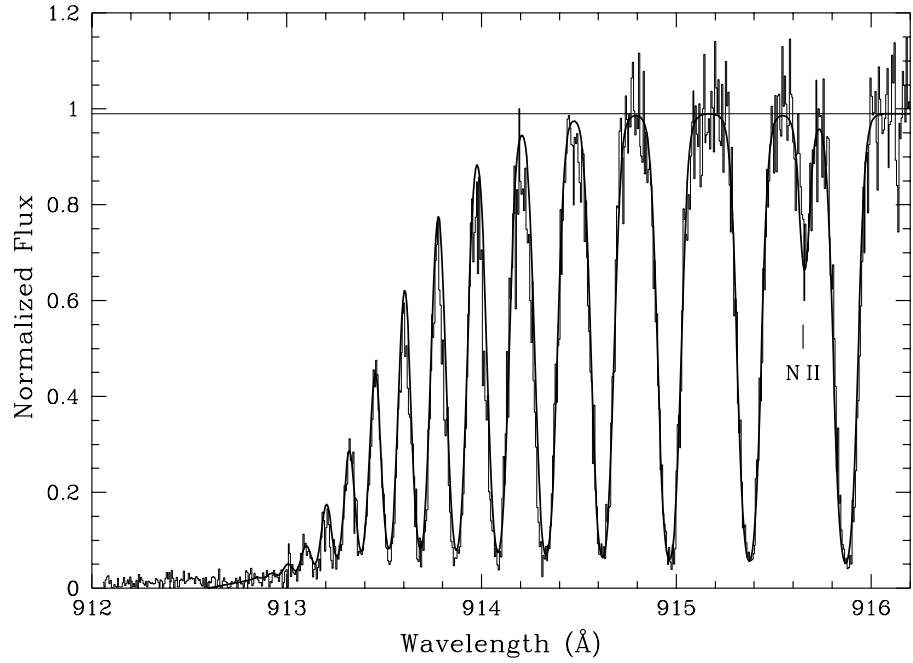}
\caption{The \fuse\ MDRS SiC1b spectrum of the converging \Hone\ Lyman 
  series is plotted as the histogrammed points, the profiles resulting
from the best fit to the \LA\ profile as the smooth line.
The assumed continuum is shown as the thin solid line.
The LWRS spectrum is essentially identical, but with lower signal to noise.
\label{fig_lyedge}   }
\end{figure}

The FUSE spectrum of the converging Lyman series might in principle
be used to try to determine the \Hone\ column.  Such fits give results
consistent with the column density derived from the \LA\ profile,
but with rather large uncertainties.  The main difficulty is that the
flux calibration is inherently uncertain at precisely the wavelengths
needed to perform the fits to the Lyman edge, because there is no
reference star available for which unattenuated continuum flux is present
at the Lyman edge.
For any given set of parameters we find that the flux is overestimated at
some wavelengths and underestimated at others, by about $\pm$5\%.
If we allow for such uncertainties we find that 
variations of the gas Doppler b parameter within \onesig\ limits 
result in variations in the derived \lognh\ of about 0.15 dex, which is
too large to provide a significant constraint on our results.
Ultimately, we may be
able to determine a more accurate calibration at these wavelengths
by means of combining data from several white dwarfs for which reliable
\Hone\ columns have been determined by other means.  The interstellar
\Hone\ absorption at the Lyman edge predicted from the column densities, 
temperatures, and turbulent velocity derived above from the GHRS data is shown 
in Figure \ref{fig_lyedge}.  The agreement is excellent, indicating 
in particular that the Doppler width of the gas is accurately known.
The effects of the broad component of the instrument LSF is clear in this
Figure, as the flux in the cores of the saturated Lyman lines is several 
times higher than the background level seen below the Lyman edge.
 
We will combine the available \euve--derived measurements of 
the \Hone\ column density 
with our GHRS / \fuse\ measurement to provide our best
estimate of \Nh.  The \euve\ measurement uncertainties are driven
in large part by the uncertainty in the effective temperature.  Each group
used different estimates for this uncertainty, but they show similar
sensitivity in the sense that a change in \Teff\ of 500K would correspond
roughly to a change in \Nh\ of $0.6 \times 10^{17}$\cmsq\ (this scaling
is not apparent in the \citealt{Barstow:1997} paper, but in the earlier
\citealt{Barstow:1995} paper \Nh\ appears to be roughly 50\% more
sensitive to changes in \Teff).  
If we take the uncertainty for \Teff\ from \citet{Dupuis:1995} as 
representative, the corresponding uncertainty in \Nh\ from an individual
\euve\ measurement is roughly equal to the \onesig\ uncertainty
in our determination of \Nh\ from the GHRS and FUSE data.
Therefore, we will simply take the mean of the four measurements (after first
averaging the two values obtained by \citet{Barstow:1997} for both the
homogeneous and stratified models), 
which gives \Nh\ = $8.511 \times 10^{17}$\cmsq, or \lognh\ = 17.930. 
The standard deviation of the four values was $0.398 \times 10^{17}$,
or 0.02 dex, but this may not be an adequate representation of the
true uncertainty.  Each of the EUV measurements is affected in a similar
fashion by the uncertainties in the effective temperature.  
\citet{Barstow:1997} were able to obtain relatively small uncertainties for
\Nh\ in the context of either the homogeneous or stratified models,
but the differences between the two corresponding column densities was
$0.5 \times 10^{17}$\cmsq, or almost as much as for the nominal
uncertainty of 500\,K for \Teff.  It is worth noting that the two
recent optical determinations of \Teff\ have uncertainties greater
than 500\,K: 49000 $\pm$ 2000\,K (\threesig) by
\citet{Napiwotzki:1993} and 50822 $\pm$ 639\,K (\onesig) by
\citet{Finley:1997}.  Similarly, as we showed above in Section
\ref{sec_los} the standard deviation for all the recent determinations
of \Teff\ is about 700\,K.  Therefore, we prefer to use a somewhat
more conservative estimate for the uncertainty in \Nh, and adopt the
value \lognh\ = 17.930 $\pm$ 0.060 (\twosig).  

\subsubsection{\Oone\ Column Density}
\label{sec-o_one}
The FUSE bandpass contains numerous \Oone\ absorption lines of varying
strength.  The total column density to \HZ\ is sufficiently low that even the
strongest \Oone\ lines are close to being on the linear 
part of the curve of growth.
The data were analyzed with profile fits and a curve-of-growth
analysis.  Good quality data were obtained in all channels for the MDRS
observation.  
Separate analyses of the day and night MDRS data actually give a somewhat higher
value for N(\Oone) than the night data; the opposite would be expected if
airglow contamination were significant.

The MDRS data were analyzed in 4 ways:  the program Owens.f was used
by two members of our group to perform independent profile fits,
and a third member of our group performed both 
a profile fit using the code of Fitzpatrick \citep{Spitzer:1993} and
a curve-of-growth analysis.  Each analysis was performed with different
choices of continuum regions to fit, different choices of polynomial order
for the continuum fits, and different combinations of background
level and LSF.  The resulting column densities, with \twosig\ 
uncertainties, were:
\logno\ = 14.47\err{0.07}{0.05}, \logno\ = 14.48\err{0.08}{0.07},
\logno\ = 14.50 $\pm$ 0.04, and \logno\ = 14.51\err{0.07}{0.06}.
The corresponding b-values and \twosig\ uncertainties from each fit were:
2.8 $\pm$ 0.2, 3, 2.5 $\pm$ 0.4, and 2.7$\pm$0.3.
The b-value for the second fit is poorly-determined because only
unsaturated lines were included, which are insensitive to b.
The first two b-values were from simultaneous fits to \Oone, \None, and \Ntwo\ 
lines in the \fuse\ data; if only the \Oone\ lines are included in the
first fit the uncertainty increases to $\pm$0.40\,\kms.
The b-values fall somewhat below the value predicted from the line of sight
velocity structure derived from the GHRS data, but are consistent
within \twosig.
The above determinations are in good agreement with one another;
we adopt the value \logno\ = 14.49 $\pm$ 0.08 (\twosig).
A sample of the fits to the \Oone\ lines is shown in Figure \ref{fig_oi_prof},
and the curve of growth is shown in Figure \ref{fig_oi_cog}. 

A number of tests were performed to examine the magnitude of the various
possible systematic effects described in Section \ref{sec_syst}.
Increasing the background level by a factor of 2 in the SiC channels
and a factor of 4 in the LiF channels increased the 
value of \logno\ by 0.02 dex, decreasing the background by a factor of 2
decreased \logno\ by $<$ 0.01 dex.
Fitting only detector 1 data or only detector 2 data, to test for
fixed-pattern noise effects and detector-dependent resolution or background
effects, changed \logno\ by
$<$ 0.01 dex.  Eliminating any single line from the fit, to test for
erroneous f-values or fixed-pattern noise effects, had no effect on \logno.
Eliminating both the line at 1039\AA\ and the multiplet at 988\AA\ from
the first profile fit causes
\logno\ to increase by 0.04 dex, but the uncertainty of the result doubles.
Eliminating these lines from the COG analysis causes the resulting \logno\ to
increase by only 0.02 dex.
The LSF was varied by doubling the amplitude of the broad component,
which increased \logno\ by 0.02 dex, and eliminating the broad component
(with a corresponding increase in the background level to match the
cores of the saturated Lyman lines), which decreased \logno\ by 0.02 dex.
These effects have all been included in the uncertainties quoted above
for the first profile fit.

%
  
\subsubsection{\None\ Column Density}
\label{sec-n_one}
As with \Oone, 
the FUSE bandpass contains numerous \None\ absorption lines of varying
strength.  In this case the low column density to \HZ\ means that many
of the lines that are ordinarily the most useful are not detected.
High quality data were obtained in all channels in the MDRS
observation, and usable data were obtained in all channels in the LWRS
observation.  
We analyzed both the complete dataset and the pure night data 
for the LWRS observation and found
that the geocoronal emission from \None\ during the day portion of
the orbit did not significantly affect the absorption line profiles.
The lines strong enough to be useful in the \fuse\ data were the 
triplet at 1134\AA\ and the lines at 953.655\AA\ and 953.970\AA.

The MDRS data were analyzed by having two members of the group perform
independent profile fits and a third member perform a curve-of-growth
fit.  The curve-of-growth analysis also included the \None\ multiplet
1199.550\AA, 1200.223\AA, and 1200.710\AA\ from the GHRS G160M
dataset.  For the MDRS data, the results were \lognn\ = 13.51
\err{0.05}{0.06},
\lognn\ = 13.51\err{0.06}{0.05}, and \lognn\ = 13.51\err{0.06}{0.05}
(\twosig\ uncertainties).
The b-value and \twosig\ uncertainty derived from the curve of growth 
analysis was 
3.0\err{1.4}{0.7}\kms; no determination of b was  possible from the
profile fits because the lines used were all on the linear part of the
curve of growth.  The adopted value for \lognn\ is 13.51 $\pm$ 0.06.
The fit to the MDRS \None\ profiles is shown
in Figure \ref{fig_ni_prof}, and the curve of growth is shown in
Figure \ref{fig_NI_cog}.

The \None\ fits were subjected to the same tests for systematic errors
as described in the previous section for \Oone.  Because all of
the \None\ lines in the \fuse\ band relatively weak, 
they are less sensitive to variations in the LSF or the
background than partially saturated lines; none of these tests caused
a variation in the \lognn\ greater than 0.01 dex.

%

\subsubsection{\Done\ Column Density}
\label{sec-d_one}

Because of the low \Hone\ column density along this line of sight, absorption by
\Done\ is expected to be detectable at \LA, \LB, and possibly at \LG.  
The \Done\ \LA\ profile is shown in Figures
\ref{fig_lya_raw} and \ref{fig_lyanorm}; it is clearly well-resolved.
The \Done\ \LB\ absorption is clearly seen in all \fuse\ channels in
the MDRS observation, but is clearly detected only in the night
portion of the LWRS observation in the LiF1 channel.  The MDRS SiC2
\LB\ spectrum was rather noisy, however, and was not included in the
fits.
\Done\ \LG\ absorption
is not detected in either SiC channel.

The GHRS \LA\ profile was first fit independent of the \fuse\ data.
The resulting value for the column density was \lognd\ = 13.16 $\pm$
0.04 (\twosig), with a b-value of 7.1 (implying the line is
well-resolved by the GHRS echelle resolution). Changing the background
level from our best estimate by $\pm$ 2\% of the continuum level
changed \lognd\ by $\pm$0.02 dex.  A simultaneous fit to the GHRS \LA\
spectrum and the \fuse\ MDRS \LB\ data from channels LiF1, LiF2, and
SiC1 gave
\lognd\ = 13.15 \err{0.04}{0.045} (\twosig).
Varying the fitting parameters for the \fuse\ data as described in
Section \ref{sec-o_one} above for \Oone\ resulted in variations in
\lognd\ of $\pm$0.01.  Profile fits to the \fuse\ MDRS data alone gave
\lognd\ = 13.11 $\pm$ 0.20.  The LWRS LiF1 \LB\ spectrum was not included in
the fits.  The \Done\ line is detected in this spectrum, and is 
consistent with the fits
to the other data, but the uncertain corrections for airglow contamination
in the LWRS spectrum, even for the night only data, means that these
data will not significantly constrain the overall fit.
The \fuse\ \Done\ line profiles are shown in Figure \ref{fig_lyb},
along with the best fit to the combined \fuse\ and GHRS datasets.

\subsubsection{Other Species}
A number of other species were detected in the ISM along
the line of sight to \HZ.  These include \Ctwo, \Cthree, \Ntwo, 
\Osix, \Sitwo, \Arone, and possibly \Nthree\ in addition
to the \Mgtwo\ and \Fetwo\ mentioned earlier.  The column densities for
these species are listed in Table \ref{tab-cogcolumns}.  Of particular
interest is the column density for \Ntwo\ (\lognntwo = 13.62\err{0.13}{0.17}),
which slightly exceeds that of
\None.  The ionization potential for \None\ is greater than that for \Hone,
so ordinarily one expects N to be predominantly neutral.  The local
ISM is dominated by hot gas that may be far from ionization
equilibrium, however, so estimates of the total abundance of nitrogen
in the local ISM must account for possible large ionization
corrections to the measured \None\ abundances.  The \Ntwo\ transitions
at 915.613\AA\ and 1083.993\AA\ are ordinarily strongly saturated, but
the total column density to \HZ\ is low enough that an accurate
\Ntwo\ column density can be determined.
The \Ntwo\ absorption lines and best-fit line profiles are shown in 
Figure \ref{fig_nii_prof}.
The \Nthree\ column density quoted in Table \ref{tab-cogcolumns}
is an upper limit because the \Nthree\ 989.799\AA\ is blended with
the \Sitwo\ 989.873\AA\ line.  If the \Sitwo\ column density derived from the
1193.290\AA\ line and the b-value derived from \Oone\ are used to predict the
equivalent width of the \Sitwo\ 989.873\AA\ line, then it is likely that at 
least half, and perhaps all, of the measured equivalent width at 989.8\AA\ is
caused by \Sitwo\ rather than \Nthree.
Even if the measured absorption results
primarily from \Nthree, the upper limit suggests that \Nthree\ is a minor
contributor to the total N column along this sightline.

No \Hmol\ was detected.  The range of possible column densities
was determined by using the profile-fitting program Owens to examine the
regions of the spectrum in the vicinity of the two-dozen strongest \Hmol\
absorption lines that were not blended with any other features.  The
column density of \Hmol\ was then gradually
increased until the change in chi-squared was unacceptable.
The resulting $3-\sigma$ upper limit to the column density of \Hmol\
was $1.2 \times 10^{13}$ \cmsq.

The column density for \Ctwo\ shown in Table \ref{tab-cogcolumns} is
larger than that for \Oone, possibly suggesting that C is over-abundant
on this sightline, that O is depleted, or that there is substantial
component of \Htwo\ gas along this line of sight.  The large column density
of \Ntwo\ supports the latter explanation.  It should
be noted also that the column density for \Ctwo\ is particularly uncertain,
because the saturation effects are large and depend on the assumption that
the b-value for \Ctwo\ is the same as that of \Oone.  If there is a
substantial contribution to the column density from \Htwo\ gas, the
effective b-value for \Ctwo\ will likely be larger than that for \Oone\ and
the \Ctwo\ column density will be overestimated accordingly.  This
source of systematic error is not included in the uncertainties quoted
in Table \ref{tab-cogcolumns}; the column density derived from the 
apparent optical depth method (log N(\Ctwo) = 13.63) is a reasonable 
lower limit.

We note that \Ctwo$^\ast$ at 1037.018 \AA\ is not detected in our 
\fuse\ observations.  The $3\sigma$ upper limit to the equivalent width
of this line in the LiF1A data is $W_\lambda \leq 5.5$ m\AA, including
a crude estimate of the continuum placement uncertainties.  This
implies $\log N(\mbox{\Ctwo}^\ast) \lesssim 12.7$ ($3\sigma$) along the
\HZ\ sight line.  If one assumes the excitation of \Ctwo$^\ast$
is due to electrons, the electron density may be calculated by
comparing the column densities of the excited- and ground-state \Ctwo:
\[
n_e \approx 0.18 \frac{N(\mbox{\Ctwo}^\ast)}{N(\mbox{\Ctwo})}T^{1/2}
\]

\citep[see][]{Spitzer:1993}.  Assuming $T=5353$ K (see Section \ref{sec-losvel})
and that $\log N(\mbox{\Ctwo}) = 14.83$ (derived using the \Oone\
$b$-value), we find $n_e \lesssim 0.1$ cm$^{-3}$.  Given the uncertainties
in the latter assumption, however, a more conservative value is $n_e
\ll 1.5$ cm$^{-3}$, derived using the lower limit for the \Ctwo\ column 
density determined with the apparent optical depth method.

\section{Summary}

The line of sight to \HZ\ was investigated using a combination of
\fuse\ spectra, archival GHRS spectra, and results obtained from
\euve\ spectra taken from the literature.  The line of sight to \HZ\
is particularly simple, with only a single velocity component
discernible.  An additional high-temperature component was found with
a very low column density, detectable only in \Hone.  The line of
sight to \HZ\ exits the LIC after a very short distance, and the
velocity of the main component of the absorbing gas is not compatible
with the velocity of the nearby G cloud in the LISM, hence this
material is presumed to be located in the North Galactic Pole cloud.
Doppler widths for each species determined from profile fits or curve
of growth analyses of the \fuse\ data were consistent with the
temperature and turbulent velocity for the gas derived from the GHRS
high resolution data.  Consistent fits to the \Done\ column density
were obtained when fitting the \fuse\ and GHRS data separately and
simultaneously, indicating that the \Done\ \LA\ line is not affected
by saturation effects.  The \fuse\ spectra were used to determine
\Oone\ and \Done\ column densities by both profile fitting and curve
of growth analyses, with consistent results obtained in each case.
Careful attention was paid to numerous potential sources of systematic
error.  Our adopted \Hone\ column density, \lognh = 17.93 $\pm$ 0.06
(\twosig), is the mean of our measurement from the GHRS \LA\ profile
and the EUV measurements of \citet{Dupuis:1995},
\citet{Barstow:1997}, and \citet{Wolff:1999}.

Our results are summarized in Table \ref{tab_summary}.  
All uncertainties in the table are \twosig\ (but values quoted below from
the literature are \onesig).
Our result for D/H along the line of sight to \HZ\ is 
$(1.66 \pm 0.28)\times 10^{-5}$.  
The GHRS dataset analyzed in this work has been previously analyzed by
\citet{Landsman:1996}, who obtained the similar value for D/H
of $1.6 \times 10^{-5}$.  
Our value for D/H along this sightline exceeds by approximately 
1.5\,$\sigma$\ both
the mean value reported by \citet{Linsky:1998} for measurements of D/H
within 100\,pc of the sun, $(1.47 \pm\ 0.10) \times 10^{-5}$, 
and the mean value of the \fuse\ measurements reported in this first set
of papers \citep{Moos:2001}.

Our result for \Oone/\Hone\ is $(3.63 \pm\,0.84)\times 10^{-4}$.
Converting this ratio to a logarithmic abundance
log~(\Oone/\Hone)~+~12.00 gives 8.56\err{0.09}{0.11}.  The ionization
balance for both O and H are linked by resonant charge exchange
reactions (see \citealt{Jenkins:2000} and references therein), and
there is no apparent \Hmol\ present, so \Oone/\Hone\ should be
representative of O/H along this line of sight.  
\citet{Meyer:1998} report an average gas-phase abundance for 13 sight lines
of $(3.43 \pm\,0.15)\times 10^{-4}$ \citep[after correction for our
preferred \Oone\ $\lambda1355$ oscillator strength of $f =
1.16\times10^{-6}$ from ][]{Welty:1999}, and they estimate no more
than O/H$\sim1.8\times 10^{-4}$ can be incorporated into grains,
implying an upper limit to the total (gas+dust) ISM oxygen abundance
of $5.2\times 10^{-4}$.  Our O/H measurement is consistent with the
average gas-phase value of \citet{Meyer:1998}.

\citet{Sofia:2001a,Sofia:2001b} have demonstrated
that the value of O/H determined from F and G star photospheres,
$(4.45 \pm 1.56)\times 10^{-4}$, is consistent with two recent
determinations of the solar abundance, $(5.45 \pm 1.00)\times 10^{-4}$
\citep{Holweger:2001} and $(4.90 \err{0.60}{0.53})\times 10^{-4}$
\citep{AllendePrieto:2001}, and with the  total oxygen
abundance estimated from the work of \citet{Meyer:1998}.
\citet{Sofia:2001a,Sofia:2001b} argue that the recent solar system
values should therefore be used as the ISM abundance standard for
oxygen.  Our value for O/H in the gas phase along the line of sight to
\HZ\ is significantly lower than each of these values, although not by
more than \twosig\ (given the large uncertainties in the solar system
determinations).  Our result is in agreement with the compilation of
B-star abundances presented by \citet{Sofia:2001a,Sofia:2001b},
$(3.50 \pm 1.33)\times 10^{-4}$, although these authors argue that
B-star abundances are unlikely to be appropriate measures of the ISM
standard.  It seems likely that the gas-phase oxygen abundance
measured by \citet{Meyer:1998} implies modest depletion of oxygen in
the ISM towards more distant stars; given the agreement of our O/H
measurement with the Meyer et al. average, it seems likely a similar
degree of oxygen depletion is present in the cloud observed along the
\HZ\ sight line.

The ionization fractions for N and H are not as strongly coupled as
those of O and H \citep{Jenkins:2000}, so it is possible for N to be
more highly ionized than H, and therefore for the gas-phase abundance
of N relative to its solar system value to be less than that of O.  If
we combine our column densities for \None, \Ntwo, and the maximum
likely column density for \Nthree\ (assuming only half of the measured
absorption is due to \Sitwo), we obtain a total nitrogen column
density of $N_N = 7.88\err{1.56}{1.43} \times 10^{13}$, and a ratio
N/\Hone\ = $(9.26 \pm 1.87)\times 10^{-5}$.  This ratio provides an
upper limit to N/H, corresponding to the limit that that hydrogen
along this line of sight is predominantly neutral.  Given the
significant column density of \ion{N}{2}, it is reasonable to expect
that there will also be a significant column density of \ion{H}{2}
present.  Converting to a logarithmic scale with \lognh\ = 12.00 gives
an abundance for N of 7.97\err{0.08}{0.10}.  This is consistent with
the solar abundance value of
\citet{Grevesse:1993}, 7.97, and the more recent value from
\citet{Holweger:2001} of 7.93 $\pm$ 0.11,
and somewhat exceeds the value of 7.81 for
nearby B stars \citep{Gies:1992}.
We can set a lower limit to the N abundance by setting the ionization
fraction of H to the upper limit of 50\% found by 
\citet{Dupuis:1995} (the other \euve\ analyses obtained lower ionization
fractions for hydrogen).  The resulting logarithmic abundance 
is 7.79\err{0.08}{0.10},
essentially equal to the value for B stars, and consistent with the
ISM gas phase abundance of \citet{Meyer:1997}.
Thus N is not significantly depleted onto
dust grains along this line of sight.  

Further discussion of the implications of these results and comparisons
with other sightlines can be found in the companion paper by \citet{Moos:2001}.

\acknowledgements
This work is based on data obtained for the Guaranteed Time Team 
by the NASA-CNES-CSA FUSE mission operated by the Johns Hopkins 
University. Financial support to U. S. participants has been 
provided by NASA contract NAS5-32985.  French participants are
supported by CNES.  We would like to thank Ed Fitzpatrick for use
of his profile fitting software.



\clearpage

\clearpage

\clearpage

\clearpage

\clearpage

\clearpage
\begin{figure}
\plotone{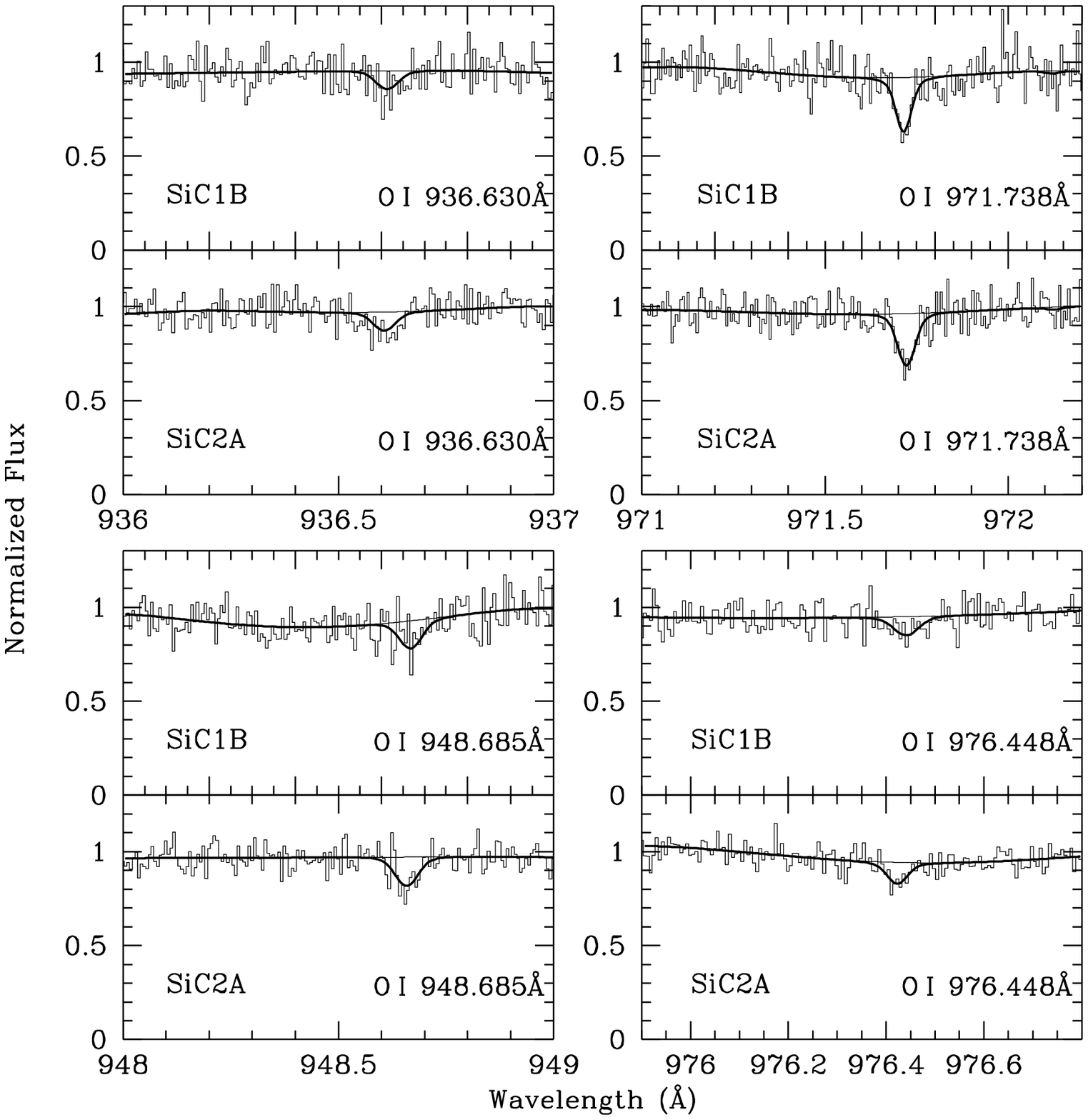}
\end{figure}

\clearpage
\begin{figure}
\plotone{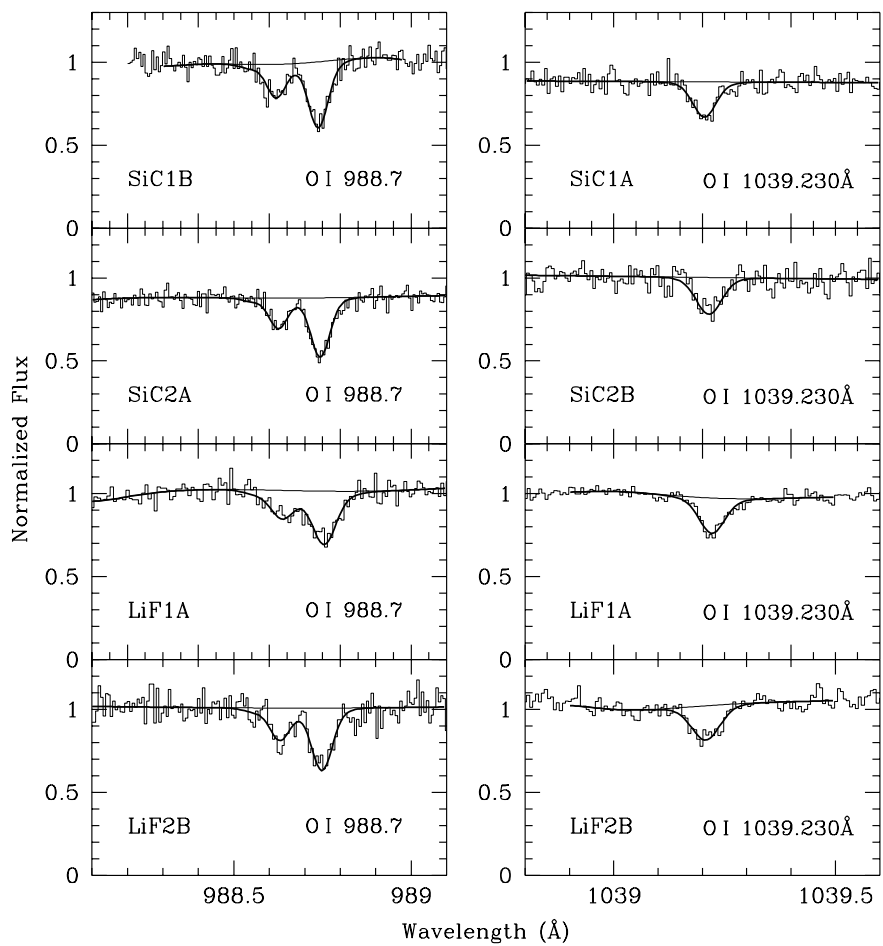}
\caption{The observed \fuse\ MDRS \Oone\ line profiles are shown as the
  histogrammed data points, the continuum fits are the thin solid lines,
  and the profiles computed from the adopted \Oone\ column density and
  doppler width is plotted as the thick solid curve. 
  The data are un-binned.
\label{fig_oi_prof}   }
\end{figure}

\begin{figure}
\plotone{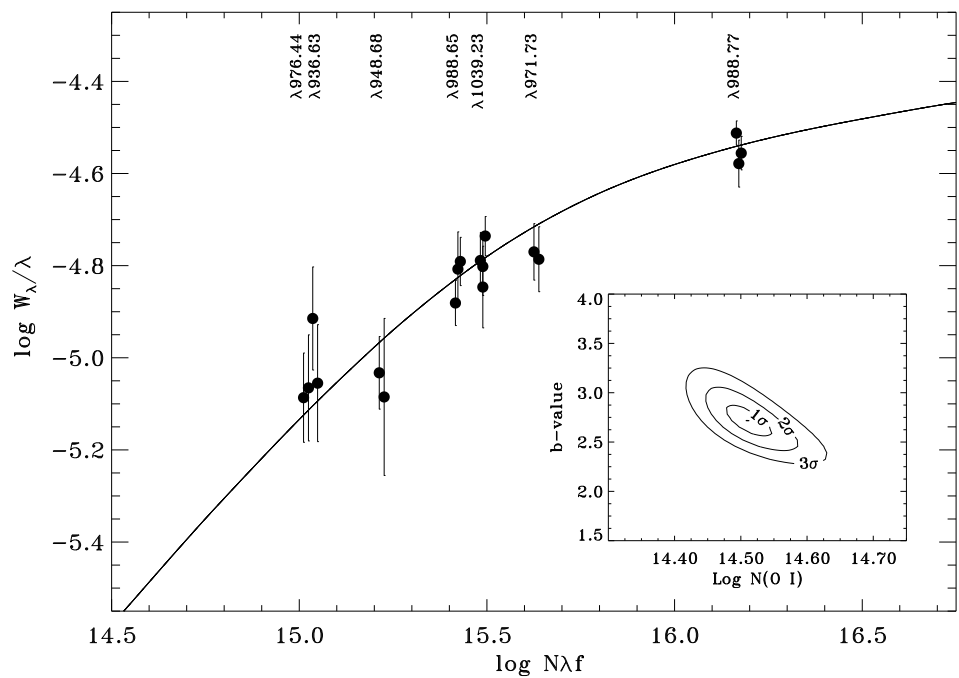}
\caption{The equivalent widths of \Oone\ lines measured
  from FUSE MDRS data are plotted, along with the best-fit curve of growth
(b = 2.72\,\kms).
The inset shows contours of constant $\Delta$\chisq\ in the b--log\,N plane.
The measurements of lines from multiple channels have been offset slightly
in log\,N$\lambda$f to avoid confusion.  The wavelengths of the lines are 
marked along the top of the figure.  The plotted errorbars indicate the
\onesig\ measurement uncertainties in the equivalent widths.
\label{fig_oi_cog}   }
\end{figure}

\clearpage
\begin{figure}
\plotone{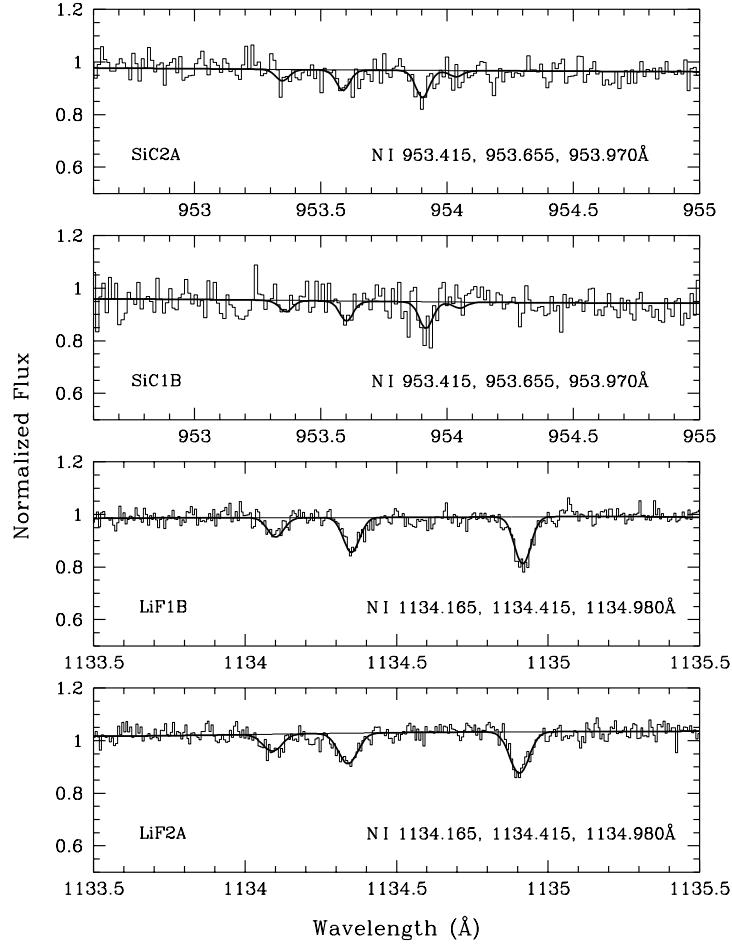}
\caption{The FUSE MDRS \None\ line profiles are shown as the thin histogram
and the best fit is shown as the thick solid line.  The continuum fit is
the thin smooth line.  The SiC data have been binned by two pixels for
plotting purposes.  
\label{fig_ni_prof}   }
\end{figure}

\clearpage
\begin{figure}
\plotone{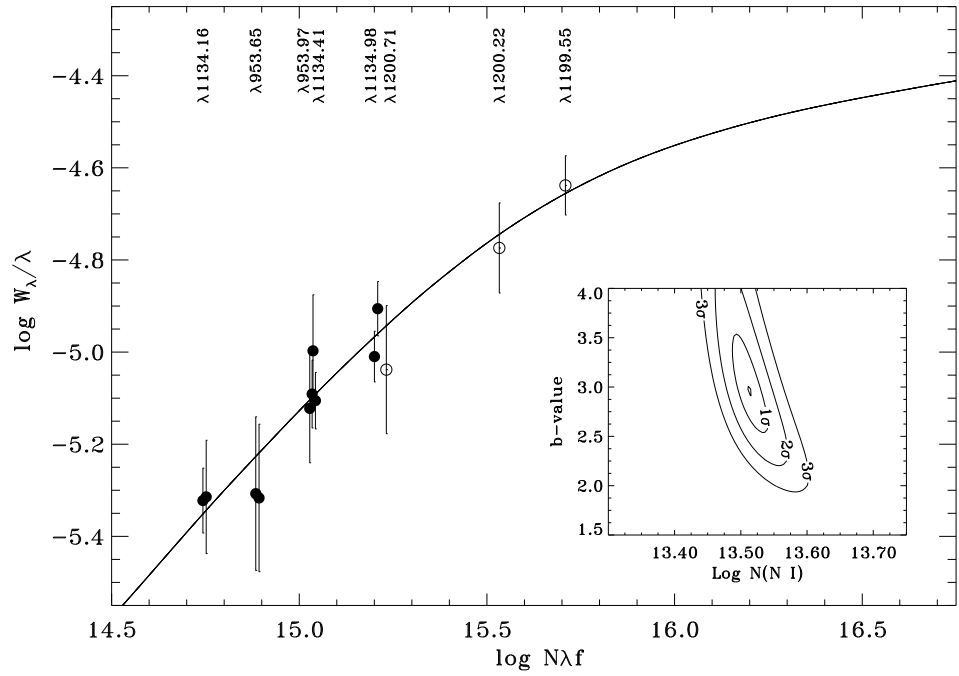}
\caption{As for Figure \ref{fig_oi_cog}, but for \None.
  The solid circles are from FUSE MDRS data, the open circles from 
intermediate-resolution GHRS spectra.
\label{fig_NI_cog}   }
\end{figure}

\clearpage

\begin{figure}
\plotone{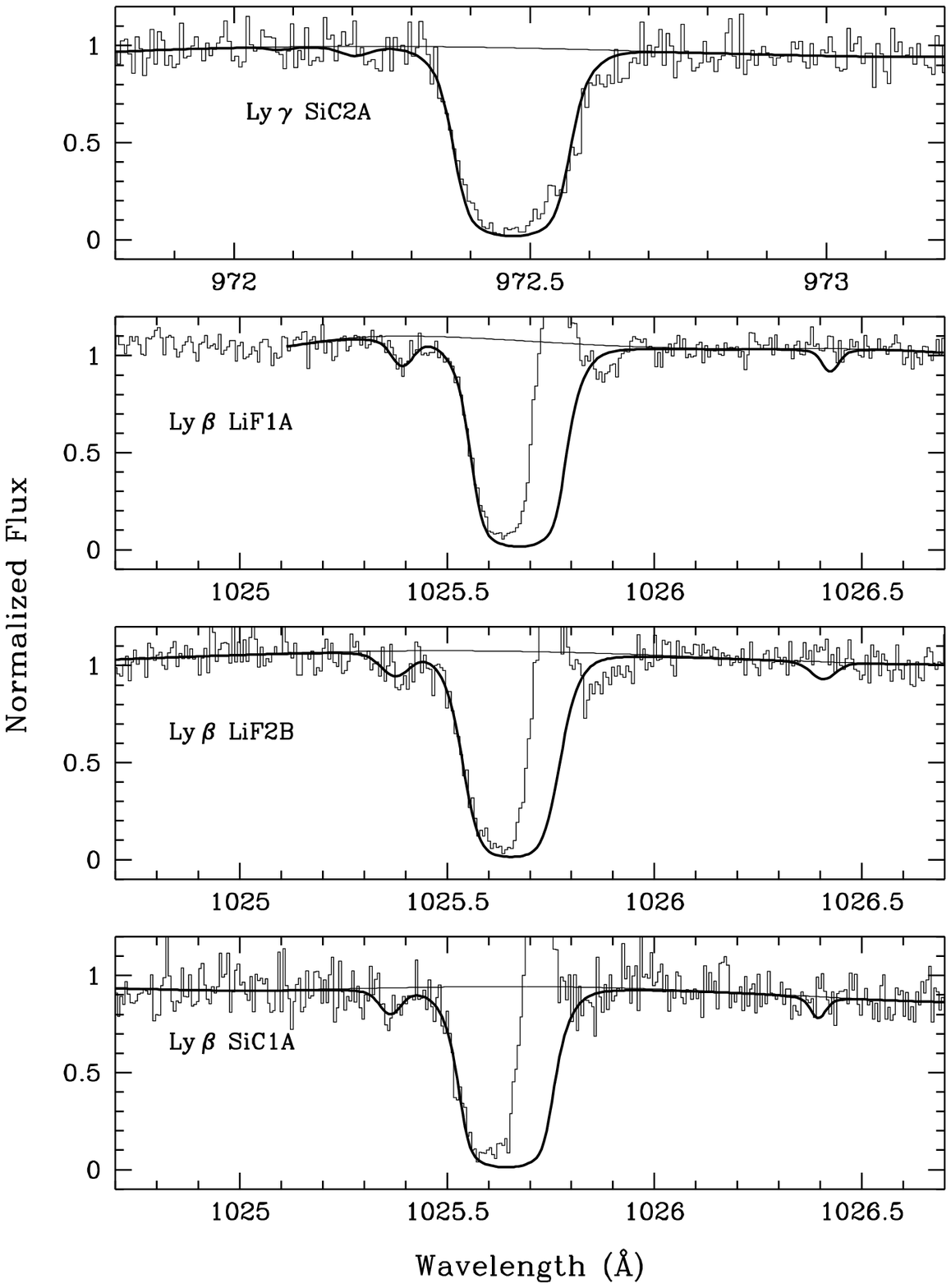}
\caption{The FUSE spectra of \Hone\ and \Done\ \LB\  and \LG\ 
  absorption are plotted as the thin histogram, and the best fit to the combined
GHRS and \fuse\ data as the thick smooth line.  No \Done\ absorption is
seen at \LG, but it is clearly seen at \LB.  
The geocoronal \LB\ emission  is clearly seen in the red
wing of the interstellar absorption; it is well-separated from \Done.
Geocoronal emission at \LG\ is present, but quite weak.
The \Oone\ 1026.473\AA\ absorption predicted from the measured \Oone\ column
density is also plotted; no absorption is seen at this wavelength so
it has been assumed that the oscillator strength for this line is in error.
It was not included in any fits.
\label{fig_lyb}   }
\end{figure}

\clearpage
\begin{figure}
\plotone{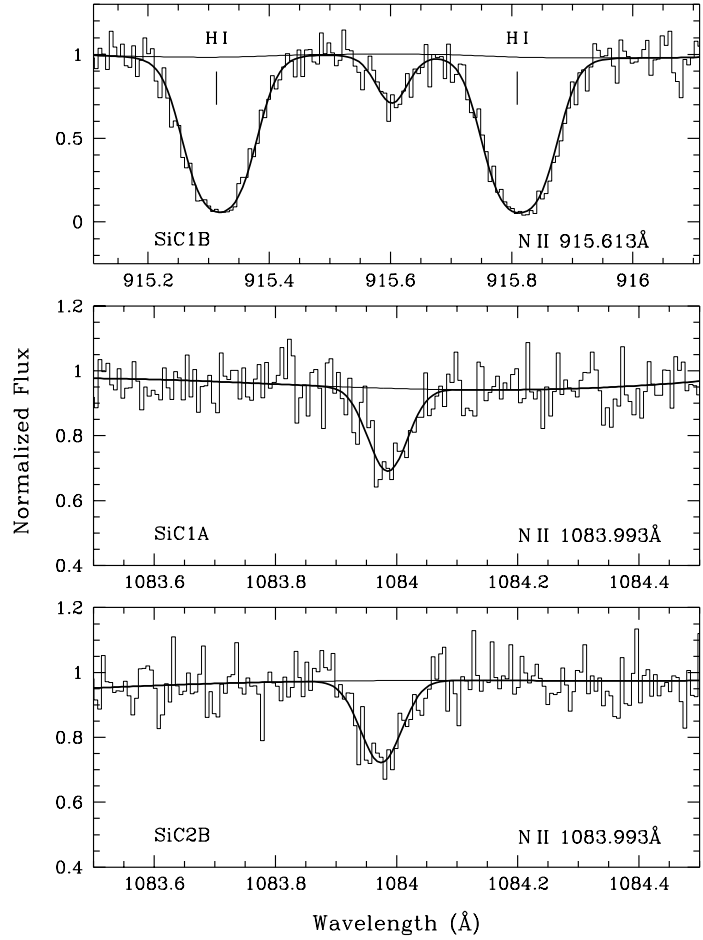}
\caption{The FUSE MDRS \Ntwo\ line profiles are shown as the thin histogram
and the best fit is shown as the thick solid line.  The continuum fit is
the thin smooth line.  The neighboring interstellar \Hone\ line profiles
are labelled with tick marks in the plot of the 915.663\AA line.
\label{fig_nii_prof}   }
\end{figure}

\clearpage
\include{tab1}

\clearpage
\include{tab2}

\clearpage
\include{tab3}

\clearpage
\include{tab4}

\clearpage

\newcommand{\bmark}{$^d$}

\include{tab5}

\clearpage

\include{tab6}

\end{document}

%% file: tab1.tex
\begin{deluxetable}{lcc}
\tablecolumns{8}
\tablewidth{0pt}
\tablecaption{Target Summary for HZ\,43A \label{tab-tgtinfo}}
\tablehead{
\colhead{Quantity} &
\colhead{Value} &
\colhead{Reference}
}
\startdata
 Spectral Type & DA1 & 1 \\
 $\alpha_{2000}$\tablenotemark{a} & $13^{h}16^{m}21\fs78$ & 2 \\
 $\delta_{2000}$\tablenotemark{a} & +29\arcdeg 05\arcmin 55\farcs5 & 2 \\
 $l$ & $54.11\degr$ & 2 \\
 $b$ & $+84.16\degr$ & 2 \\
 $d$\tablenotemark{b} \ (pc)& $68\pm 13$ & 3 \\
 $V$ & 12.914 & 2 \\
$B-V$ & $-0.31$ & 2\\
$T_{\rm eff}$\ (K)& 50,900 & 1 \\
$\log g$\ (cm s$^{-2}$) & 8.0 & 1 \\
\enddata
\tablenotetext{a}{Corrected for proper motion to epoch 2000.5}
\tablenotetext{b}{Trigonometric parallax}
\tablerefs{(1)\ \citealt{Finley:1997}; (2)\ \citealt{Bohlin:1995};
(3)\citealt{vanAltena:1995}.}
\end{deluxetable}

%% file: tab2.tex
\begin{deluxetable}{cccccc}
\tabletypesize{\scriptsize}
\tablecaption{Log of \fuse\ observations. \label{tab-fuseobs}}
\tablewidth{0pt}
\tablehead{
\colhead{Dataset ID} & \colhead{Aperture} & \colhead{Obs date} & 
\colhead{$T_{exp}$ (total)\tablenotemark{a}} & 
\colhead{$T_{exp}$ (night)\tablenotemark{a}} & 
\colhead{$N_{exp}$\tablenotemark{b}}
}
\startdata
M1010501 & LWRS &  Feb 19 2000 & 6092 & 2333 & 8 \\
P1042301 & LWRS &  Apr 22 2000 & 14447 & 5136 & 23 \\
P1042302 & MDRS &  Feb 08 2001 & 39606 & 20572 & 85 \\
\enddata
\tablenotetext{a}{Exposure duration in seconds.}
\tablenotetext{b}{Number of individual exposures.}
\end{deluxetable}

%% file: tab3.tex
\begin{deluxetable}{cccccc}
\tablecolumns{6}
\tabletypesize{\scriptsize}
\tablecaption{Log of GHRS observations. \label{tab-ghrsobs}}
\tablewidth{0pt}
\tablehead{
\colhead{Dataset} & \colhead{Grating} & \colhead{Aperture and} & 
\colhead{Wavelength range} & \colhead{Start} & \colhead{Exposure} \\
\colhead{ID} & \colhead{and order} & \colhead{substep pattern} & 
\colhead{\AA} & \colhead{Time (UT)\tablenotemark{a}} &
\colhead{Time (s)} \\
}
\startdata
Z2R50104T & Ech A - 46 & LSA 7 & 1212-1218 & 00:21 & 7834 \\
Z2R50107T & Ech A - 46 & LSA 7 & 1212-1218 & 05:05 & 7181 \\
Z2R50204T & Ech B - 24 & LSA 7 & 2376-2387 & 09:59 & 2176 \\
Z2R50205T & Ech B - 20 & LSA 7 & 2791-2807 & 11:27 & 2176 \\
Z2R50209T & G160M & SSA 5 & 1192-1228 & 13:10 & 1414 \\
\enddata
\tablenotetext{a}{All exposures began on July 30 1995.}
\end{deluxetable}

%% file: tab4.tex
\begin{deluxetable}{lrccrc}
\tablecolumns{6}
\tablewidth{0pc}
\tablecaption{Equivalent Width Measurements
        \label{tab-eqwidths}}
\tablehead{
\colhead{Species} & 
\colhead{$\lambda_c$\tablenotemark{a}} & 
\colhead{$\log \lambda f$\tablenotemark{b}} & 
\colhead{Channel} &
\colhead{$W_\lambda$ [m\AA]\tablenotemark{c}} &
\colhead{S/N\tablenotemark{d}}
}
\startdata
\cutinhead{FUSE}
\ion{C}{2} & 1036.337 	& 2.088 & LiF1A & 37.8\err{1.7}{2.7} & 42 \\
		& &		& LiF2B & 39.8\err{2.2}{2.8} & 26 \\
		& &		& SiC1A & 39.4\err{2.2}{2.4} & 19 \\
\\
\ion{C}{3} & 977.020	& 2.870 & SiC1B	& 14.6\err{3.1}{2.2} & 16 \\
		& &		& SiC2A & 15.6\err{1.7}{1.5} & 22 \\
\\
\ion{N}{1} & 953.655	& 1.376	& SiC1B &  4.6\err{2.2}{1.9} & 15 \\
	 	& & 		& SiC2A &  4.7\err{2.3}{2.1} & 17 \\
\ion{N}{1} & 953.970	& 1.521	& SiC1B &  9.6\err{3.9}{2.3} & 16 \\
	 	& & 		& SiC2A &  7.2\err{1.9}{2.6} & 20 \\
\ion{N}{1} & 1134.165	& 1.238	& LiF1B &  5.5\err{1.7}{1.9} & 47 \\
	 	& & 		& LiF2A &  5.4\err{1.0}{0.9} & 47 \\
\ion{N}{1} & 1134.415	& 1.528	& LiF1B &  8.9\err{1.7}{1.0} & 46 \\
	 	& & 		& LiF2A &  9.2\err{1.5}{1.9} & 43 \\
\ion{N}{1} & 1134.980	& 1.693	& LiF1B & 14.1\err{1.6}{2.5} & 46 \\
	 	& & 		& LiF2A & 11.1\err{1.2}{1.8} & 43 \\
\\
\ion{N}{2} &  915.613	& 2.180 & SiC1B & 20.3\err{4.7}{3.4} & 13 \\
\ion{N}{2} & 1083.994	& 2.097 & SiC2B & 23.3\err{3.0}{3.1} & 14 \\
\\
\ion{N}{3} &  989.799  	& 2.085 & SiC2A &  
	$<8.2$\err{2.9}{1.7}\tablenotemark{e} & 25 \\
\\
\ion{O}{1} & 936.630	& 0.534	& SiC1B &  8.3\err{2.9}{2.7} & 13 \\
		& &		& SiC2A & 11.4\err{4.6}{2.1} & 15 \\
\ion{O}{1} & 948.685	& 0.778	& SiC1B &  7.8\err{4.8}{2.7} & 12 \\
		& &		& SiC2A &  8.8\err{1.9}{1.6} & 15 \\
\ion{O}{1} & 971.738	& 1.123\tablenotemark{f}	
				& SiC1B & 15.9\err{2.6}{3.0} & 10 \\
		& &		& SiC2A & 16.5\err{2.5}{2.5} & 13 \\
\ion{O}{1} & 976.448	& 0.509 & SiC1B &  8.4\err{2.7}{2.4} & 15 \\
		& &		& SiC2A &  8.0\err{2.0}{2.0} & 18 \\
\ion{O}{1} & 988.655	& 0.914	& SiC1B & 16.0\err{2.2}{1.9} & 20 \\
		& &		& SiC2A & 13.0\err{1.7}{1.4} & 23 \\
		& &		& LiF2B & 15.4\err{3.6}{2.7} & 14 \\
\ion{O}{1} & 988.773	& 1.662	& SiC1B & 27.5\err{2.1}{2.7} & 20 \\
		& &		& SiC2A & 30.4\err{1.6}{2.2} & 24 \\
		& &		& LiF2B & 26.1\err{3.5}{3.0} & 14 \\
\ion{O}{1} & 1039.230	& 0.980	& LiF1A & 19.1\err{1.7}{2.2} & 42 \\
		& &		& LiF2B & 16.9\err{2.9}{2.2} & 26 \\
		& &		& SiC1A & 16.4\err{2.2}{2.9} & 19 \\
\\
\ion{O}{6} & 1031.926 	& 2.137 & LiF1A & 8.5\err{2.1}{1.7} & 39 \\
		& &		& LiF2B & 8.4\err{2.7}{2.1} & 25 \\
\\
\ion{Ar}{1} & 1048.220 	& 2.442 & LiF1A & 2.2\err{1.2}{0.8} & 49 \\
		& &		& LiF2B & $<2.4$ ($2\sigma$) & 25 \\
\cutinhead{GHRS}
\ion{N}{1} & 1199.550	& 2.192	& G160M & 27.6\err{4.2}{4.6} & 17 \\
\ion{N}{1} & 1200.223	& 2.015	& G160M & 20.2\err{5.8}{4.4} & 17 \\
\ion{N}{1} & 1200.710	& 1.713	& G160M & 11.0\err{4.3}{4.0} & 16 \\
\\
\ion{Mg}{2} & 2796.352 & 3.236 & Ech-A & 55\err{3}{3} & 12 \\
\ion{Mg}{2} & 2803.531 & 2.933 & Ech-A & 37.9\err{2.2}{2.2} & 13 \\
\\
\ion{Si}{2} & 1193.290 	& 2.775 & G160M & 24.7\err{7.0}{4.9} & 16 \\
\\
\ion{Fe}{2} & 2382.765 & 2.882 & Ech-A & 20.5\err{2.3}{2.3} & 12 \\
\enddata
\tablenotetext{a}{Central wavelength of each transition from Morton 2001.}
\tablenotetext{b}{Adopted product of wavelength and oscillator strength 
	for each transition, taken from Morton 2001.}
\tablenotetext{c}{Measured equivalent width and $1\sigma$ uncertainties 
        (in m\AA).  The uncertainties contain contributions from both
        statistical and systematic error sources.}
\tablenotetext{d}{Empirically estimated signal-to-noise ratios 
	(per unbinned pixel) for 
        continuum regions near the absorption lines of interest.}
\tablenotetext{e}{The \ion{N}{3} absorption contains a substantial 
	contribution from \ion{Si}{2} at 989.873 ($\approx +21$ \kms\
	from the center of the \ion{N}{3} line).  The equivalent width
	quoted may be dominated by the latter, but represents a firm
	upper limit for the amount of \ion{N}{3} along the HZ 43A
	sight line.}
\tablenotetext{f}{This transition is a blend of three \ion{O}{1} lines, 
	the sum of whose oscillator strengths are given here.}
\end{deluxetable}

%% file: tab5.tex
\begin{deluxetable}{llccr}
\tablecolumns{5}
\tablewidth{0pc}
\tablecaption{Curve of Growth Column Densities
        \label{tab-cogcolumns}}
\tablehead{
\colhead{Species} & 
\colhead{$\log N \pm 2\sigma$\tablenotemark{a}} &
\colhead{$b \pm 2\sigma$\tablenotemark{b}} &
\colhead{$\Delta \log N$\tablenotemark{c}}
}
\startdata
\ion{C}{2}  & 14.83\err{0.17}{0.34} & \bmark,\tablenotemark{e}  & +1.20 \\
\ion{C}{3}  & 12.59\err{0.09}{0.10} & \bmark & +0.15 \\
\ion{N}{1}  & 13.51\err{0.06}{0.05} & 3.0\err{1.4}{0.7} & \nodata \\
\ion{N}{2}  & 13.62\err{0.13}{0.17} & \bmark & +0.32 \\
\ion{N}{3}  &
	    $<12.98$\err{0.22}{0.25} & \bmark & +0.07 \\
\ion{O}{1}  & 14.51\err{0.07}{0.06} & 2.72\err{0.33}{0.26} & \nodata \\
\ion{O}{6}  & 12.91\err{0.08}{0.08} & $\infty$\tablenotemark{f} 
		& \nodata \\
\ion{Si}{2} & 12.85\err{0.18}{0.22} & \bmark & +0.21 \\
\ion{Ar}{1} & 11.96\err{0.30}{0.50} & \bmark & +0.00 \\
\ion{Mg}{2} & 12.41$\pm$0.02 & 2.8$\pm$0.3 & \nodata \\
\ion{Fe}{2} & 12.17$\pm$0.04 & 2.78$\pm$1.15 & \nodata \\
\enddata
\tablenotetext{a}{Log of column density derived from a curve of growth
	fit, with estimated $2\sigma$ 
	uncertainties.  For most species not enough transitions were
	observed to independently fit a curve of growth and determine
	a $b$-value.  In these cases we have adopted the $b$-value of
	2.72\err{0.33}{0.26} ($2\sigma$) from \ion{O}{1}.}
\tablenotetext{b}{Doppler parameter in \kms\ with estimated $2\sigma$ 
	uncertainties.}
\tablenotetext{c}{For those species for which we have adopted the 
	$b$-value from \ion{O}{1} we give the amount by which the
	curve-of-growth-corrected column density given in column 2 of this
	table exceeds the average integrated apparent column density.}
\tablenotetext{d}{We have adopted the $b$-value of
	2.72\err{0.33}{0.26} ($2\sigma$) from \ion{O}{1} to estimate
	the total column density of these species.}
\tablenotetext{e}{For \ion{C}{2} the b-value is probably larger, and log$N$
	correspondingly smaller.  Column 4 gives the maximum correction for
	this effect.}
\tablenotetext{f}{For \ion{O}{6} we assume that the instrument 
	fully-resolves the absorption profile and adopt the weighted
	average of the apparent column density integrations from the
	two LiF channels.}
\end{deluxetable}

%% file: tab6.tex
\begin{deluxetable}{ccccc}
\tablecolumns{5}
\tablewidth{0pt}
\tablecaption{HZ\,43 Column Densities and Ratios\tablenotemark{a}
 \label{tab_summary}}
\tablehead{
\colhead{ } & \colhead{\Hone} & \colhead{\Done} & \colhead{\Oone} &
\colhead{\None}
}
\startdata
Log N & 17.93 $\pm$ 0.06  & 13.15\err{0.040}{0.045} & 
14.49 $\pm$ 0.08 & 13.51 $\pm$ 0.06 \\
\cutinhead{Ratios}
\Done/\Hone & \Oone/\Hone & \None/\Hone & \Done/\Oone & \Done/\None \\
$1.66 \pm 0.28 \times 10^{-5}$ & 
$3.63 \pm 0.84 \times 10^{-4}$ &
$3.80 \pm 0.74 \times 10^{-5}$ & 
$4.57 \pm 0.96 \times 10^{-2}$ &
$4.37 \pm 0.74 \times 10^{-1}$ \\
\enddata
\tablenotetext{a}{all uncertainties are 2\,$\sigma$}
\end{deluxetable}